\documentclass[pageno]{jpaper}

\usepackage[normalem]{ulem}
\usepackage{balance}
\usepackage{graphicx}

\usepackage{amssymb}

\usepackage{algorithm}  
\usepackage{amsmath,amssymb,amsfonts}
\usepackage{bm}
\usepackage{algorithmicx}  
\usepackage{amsmath}
\usepackage{algpseudocode} 
\usepackage{hyperref}
\usepackage{wrapfig}
\usepackage{floatflt}
\usepackage{makecell}
\usepackage{wrapfig}
\usepackage{oplotsymbl}
\algtext*{EndWhile}
\algtext*{EndIf}

\algtext*{EndFor}
\algtext*{EndFunction}
\usepackage{xcolor}
\usepackage{enumitem}
\usepackage{lipsum}
\usepackage{amsmath}
\usepackage{bbm}
\usepackage{tikz}
\usepackage{xcolor}
\newcommand*\circled[1]{\tikz[baseline=(char.base)]{
            \node[shape=circle,fill,inner sep=0.7pt] (char) {\textcolor{white}{#1}};}}
\newcommand\blankfootnote[1]{%
            \let\thefootnote\relax\footnotetext{#1}%
            \let\thefootnote\svthefootnote%
          }

\newcommand{\Fig}[1]{Fig.~\ref{#1}}

\newcommand{\Tbl}[1]{Tbl.~\ref{#1}}
\newcommand{\Sec}[1]{Sec.~\ref{#1}}
\newcommand{\Equ}[1]{Equ.~\ref{#1}}
\newcommand{\Alg}[1]{Alg.~\ref{#1}}

\newcommand{\proj}{\textsc{Juno}}

\newcommand{\revision}[2]{#2} 
\newcommand{\postrevision}[2]{#2}

\begin{document}
\title{\proj{}: Optimizing High-Dimensional Approximate Nearest Neighbour Search with Sparsity-Aware Algorithm and Ray-Tracing Core Mapping}
\date{}
\author{Zihan Liu$^{1,2}$, Wentao Ni$^{1}$, Jingwen Leng*$^{1,2}$, Yu Feng$^{3}$\\Cong Guo$^{1,2}$, Quan Chen$^{1,2}$, Chao Li$^{1,2}$, Minyi Guo*$^{1,2}$ and Yuhao Zhu$^{3}$\\
$^{1}$ Shanghai Jiao Tong University, $^{2}$ Shanghai Qi Zhi Institute, $^{3}$ University of Rochester\\
$\lbrace$altair.liu, wennitao, guocong$\rbrace$@sjtu.edu.cn, $\lbrace$ leng-jw, chen-quan, lichao, guo-my$\rbrace$@cs.sjtu.edu.cn\\yfeng28@ur.rochester.edu, yzhu@rochester.edu}

\maketitle

\begin{abstract}


    
Approximate nearest neighbor (ANN) search is a widely applied technique in modern intelligent applications, such as recommendation systems and vector databases. Therefore, efficient and high-throughput execution of ANN search has become increasingly important. In this paper, we first characterize the state-of-the-art product quantization-based method of ANN search and identify a significant source of inefficiency in the form of unnecessary pairwise distance calculations and accumulations. To improve efficiency, we propose \proj{}, an end-to-end ANN search system that adopts a carefully designed sparsity- and locality-aware search algorithm. We also present an efficient hardware mapping that utilizes ray tracing cores in modern GPUs with pipelined execution on tensor cores to execute our sparsity-aware ANN search algorithm. Our evaluations on four datasets from 1 to 100 million search points demonstrate 2.2$\times$-8.5$\times$ improvements in search throughput. Moreover, our algorithmic enhancements alone achieve a maximal 2.6$\times$ improvement on the hardware without the acceleration of the RT core.

\end{abstract}
\section{Introduction}
\label{sec:Introduction}
\blankfootnote{* Jingwen Leng and Minyi Guo are corresponding authors of this paper.}
Computer applications are becoming more intelligent with the breakthrough from deep learning~\cite{DLRM,DCN}.
One of the key data structures in these applications is high-dimensional embedding vector~\cite{KNNDiffusion, ROI, PointNet, Attention, PointConv}, which are usually generated from some kind of transformation or learning to the raw data like text, images, audio, video, and others.
Using embeddings allows for fast and accurate nearest neighbor (NN) search and retrieval of data based on their distance.
For example, we can perform the NN search to find images similar to a given image based on their content and style.

Embedding vectors are often high-dimensional, ranging from tens to thousands.
The curse of dimension makes the exact NN search computationally expensive so that approximate nearest neighbor (ANN) search becomes increasingly popular in industrial practice~\cite{ANNA,CVPRANN,ANNSurvey,LinearANN,RecommendationANN}.
ANN search trades search quality (measured in recall) for search performance (measured in throughput) in different scenarios, with GPUs being widely used to achieve higher throughput.


The \texttt{IVFPQ} technique, which combines the inverted file index (i.e., \texttt{IVF}) with product quantization (i.e., \texttt{PQ}), is the most commonly used ANN method~\cite{ANNA, FAISS}.
This method is often combined with other techniques to improve performance~\cite{malkov2018efficient}. The \texttt{PQ}-based approach encodes search point projections using a codebook in every subspace offline and then searches for the nearest neighbors by accumulating the distance information distributed in these subspaces online. However, this process requires a significant number of pairwise distance calculations between entry and query projections in low-dimensional subspaces, as well as frequent look-ups to calculate the total distance for every query.


In this work, we use the state-of-the-art ANN framework FAISS~\cite{FAISS} to study the efficiency of the \texttt{IVFPQ} method.
Although the framework employs hundreds of codebook entries to encode search points in each subspace, only a fraction of these entries is used by the top-100 results returned by a query. Notably, in some subspaces, all top-100 search points are encoded with just one entry. This entry-level sparsity provides opportunities to avoid calculating the pairwise distances of unused entries and to skip the distance look-up and accumulation of search point projections encoded by these unused entries.

In addition, our findings suggest that entries with high usage frequency demonstrate a significant degree of spatial locality, rendering the exploitation of sparsity more advantageous. Despite being sparsely distributed in memory, these entries are closely positioned in Euclidean space. In some situations, by selecting the 25\% closest entries, we can obtain all the used entries in response to a query within a subspace. Therefore, only entries that are close to the query projections are essential.


We propose a selective codebook construction algorithm to exploit the above sparsity and spatial locality to accelerate high-dimensional ANN search. 
Our approach involves using an adaptive distance threshold in each subspace to only select necessary entries. We have identified a strong correlation between the distance threshold and search point density, and we exploit it by training a simple regression model offline, with the density serving as the input. During runtime, we utilize the threshold value determined by the regression model to choose a small fraction of the search points.



Our sparsity-aware ANN method heavily uses the distance comparison operation, which is well suited for ray tracing (RT) cores in modern GPUs~\cite{Turing, Ampere, Ada}.
The RT cores implement the tree-based intersection check~\cite{kdtree, octree}, which has a reduced logarithmic time complexity.
As a result, RT cores can find objects intersected with a given ray with high efficiency~\cite{OptiX}.
Obviously, this is quite similar to our algorithmic intuition that finds close entries for a query projection in every subspace. 
Intuitively, we can organize entries as objects and query projection as rays in every subspace and let the RT core efficiently find the intersected entries of a query ray.




We present \proj{}, a fast and high-throughput ANN search system that employs algorithmic enhancement and RT core mapping to leverage sparsity and spatial locality. However, utilizing the RT core in an enhanced ANN algorithm still poses unique challenges. Firstly, calculating distances for selected codebook entries after filtering is still necessary. Secondly, naively implementing adaptive dynamic radius in RT cores would cause unacceptable scene preparation overhead during runtime. To overcome these issues, we exploit the concept of "time" in the ray tracing scenario. Specifically, we utilize the hit time results from RT cores to efficiently calculate distances, thereby avoiding costly global memory access. Additionally, we convert the dynamic distance threshold to a dynamic maximum travel time for rays, thereby avoiding online scene preparation. 
Finally, we optimize \proj{} to support inner product similarity and efficient RT-Tensor core pipelining~\cite{TensorCore}.

We evaluate \proj{} on multiple datasets sizing from 1M to 100M with both L2 distance and inner product similarity. Together they deliver an average (a maximum) of $4.4\times$ ($8.5\times$) and $2.1\times$ ($3.2\times$) improvement in search throughput in low and high search quality against the baseline. Moreover, the improvement is bound by the performance of RT cores. 
We make the following main contributions in this work:
\begin{itemize}
    \item We study the inefficiency of the typical \texttt{IVFPQ} pipeline and identify sparsity and spatial similarity in codebook usage.\vspace*{0.1cm}
    \item We design a threshold-based selective algorithm to rapidly filter out the unnecessary search points leveraging the sparsity and spatial locality and propose a mapping for our algorithm to run on the RT core.\vspace*{0.1cm}
    \item \revision{R-Q}{To the best of our knowledge, we are the first to study how to generalize the existing kNN-RT core mapping to  ANN search with arbitrary dimensions, in aspects of approximation method, metrics and system design. Based on our experimental analysis and insights, we propose \proj{}, an end-to-end high-dimensional ANN search engine with both algorithmic enhancement and optimized hardware mapping.}\vspace*{0.1cm}
    \item We quantify the effectiveness of \proj{} over existing ANN framework FAISS~\cite{FAISS}, with detailed breakdown analysis.
\end{itemize}

\section{Background}
\label{sec:Background}
This section introduces the typical process of ANN search, ray tracing pipeline, and its application in 2D/3D ANN search.

\begin{table}[h]
  \caption{Notations used in this work.}
  \label{tab-nota}
  \centering
  \begin{tabular}{|c|c|c|c|}
    \hline
    $N$ & \# search points & $D$ & search points dimension \\
    \hline
    $C$ & \# clusters & $E$ & \# codebook entries \\
    \hline
    $x$ & search point vector & $M$ & subspaces dimension\\
    \hline
    $q$ & query vector & $r$ & bounding radius\\
    \hline
    $s$ & subspace id & $e$ & codebook entry id\\
    \hline
    \multicolumn{2}{|c|}{$nprobs$} & \multicolumn{2}{c|}{\# chosen clusters in \emph{filtering}}\\
    \hline
  \end{tabular}
\end{table}

\subsection{Approximate Nearest Neighbor (ANN) Search}
\label{subsec:ann_background}


The objective of the nearest neighbor (NN) search is to identify the top-k most similar points to a query within a given set.
Typically, the similarity between two points is determined using L2 distance or inner (or dot) product, as illustrated in \Equ{eq:metric}. 
L2 distance is lower-is-better and widely used in measuring image similarity~\cite{pami2005}.
Inner product is higher-is-better and frequently used in large language models (LLMs) and Transformer architectures~\cite{GPT4, T5}.
\vspace{-0.2cm}
\begin{equation*}
\label{eq:metric}
\begin{aligned}
L2(\bm{q},\bm{x})=\sum_{i=0}^{D-1}(x_i-q_i)^2,~~~
IP(\bm{q},\bm{x})=\sum_{i=0}^{D-1}x_i\cdot q_i
\end{aligned}
\end{equation*}

\vspace{-0.2cm}
The exact NN search is often expensive and thus slow. 
Many practical cases can tolerate certain search inaccuracies, which can be exploited to improve the search throughput.
This is referred to as approximate NN (ANN) search~\cite{ANN1999}.
One of the most popular ANN techniques is inverted file index with product quantization (\texttt{IVFPQ}), which is used by top-performing ANN frameworks such as FAISS~\cite{FAISS} and ScANN~\cite{ScANN}. Notices that there are other indexing and encoding techniques, \postrevision{PR-A}{and we will discuss their details in \Sec{sec:RelatedWorks}.}

\begin{figure}[t]
    \centering
      \includegraphics[width=0.99\linewidth]{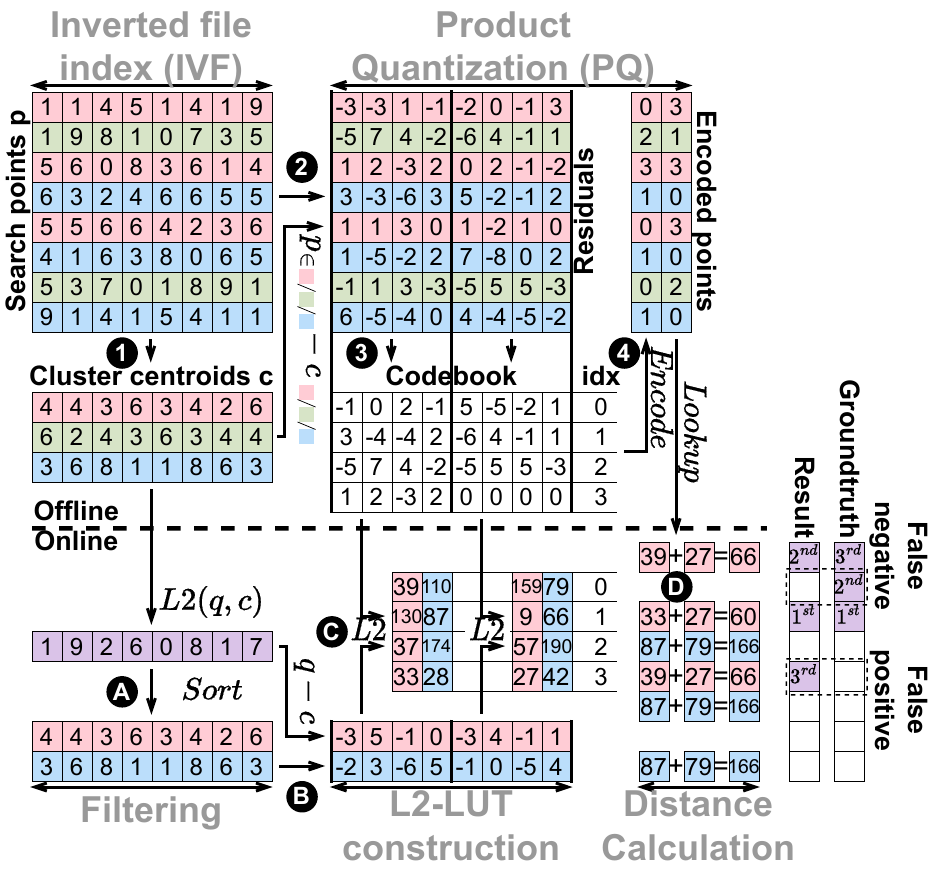}
    \caption{The example of offline training (top) and online searching (bottom) component of \texttt{IVFPQ}-based ANN search.}
    \label{fig-Example} 

\end{figure}

\Fig{fig-Example} shows an example of the \texttt{IVFPQ} technique, with notations defined in \Tbl{tab-nota}, which has an offline (top) and an online component (bottom).
The offline component relies on the inverted file index (\texttt{IVF}) and product quantization (\texttt{PQ}), and the online component consists of \emph{filtering}, \emph{L2-LUT construction} and \emph{distance calculation}, as described as follows.

 
\paragraph{Inverted File Index (\texttt{IVF})}
\circled{1} Given a query point $q$, the \texttt{IVFPQ} technique performs the first coarse-grained filtering step to identify a set of candidates from all search points.
This step is commonly implemented through \texttt{k-means}, which generates $C$ clusters on $N$ search points of full dimension $D$.
The inverted file index (\texttt{IVF}) stores the associated search points for each cluster centroid.
This step calculates the distance between query point $q$ and each centroid and identifies the closest centroid(s).
The \texttt{IVF} data structure lets us quickly locate the associated search points for selected centroid(s).


\paragraph{Product Quantization (\texttt{PQ})}
The \texttt{PQ} method is widely used for vector compression. \circled{2} Initially, the $D$-dimensional space is divided into $D/M$ $M$-dimensional subspaces. 
\revision{R-A}{\circled{3} Next, in each subspace of total $D/M$ subspaces, $E$ clusters are generated with projections of residuals (in this specific subspace) between search points and first cluster centroids. }
We refer to these $E$ clusters as the "second" clusters.
The centroids from the second clusters are combined to form the codebook, with each centroid serving as a codebook entry. 
\circled{4} Finally, search points are encoded using this codebook by replacing the projection on a subspace with the cluster ID to which the residual projection belongs. \texttt{PQ} reduces the storage space required for each search point originally in float format from $D \times \text{sizeof(float)} \times 8$ bits to $(D/M) \times \log_{2}E$ bits.


\paragraph{Online Search.}
\label{sec:IVFPQ}
After the \texttt{IVF} and the \texttt{PQ} are trained, the online search process begins. \circled{A} When a query arrives, it first calculates the pairwise distances with $C$ cluster centroids in \texttt{IVF}. The $nprobs$ closest centroids and their corresponding clusters are chosen. Subsequent processes only occur on the search points that belong to selected clusters. 
We refer to this initial stage as the \emph{filtering stage} following previous literature~\cite{ANNA}, as illustrated in the bottom left of \Fig{fig-Example}.


\circled{B} Next, the query calculates residuals with the $nprobs$ selected cluster centroids. 
\circled{C} For each residual, $E$ pairwise distances are computed between every codebook entry within each subspace. 
These $nprobs \times E \times (D/M)$ pairwise distances are then organized into a look-up table (LUT). 
we refer to this second stage as \emph{L2-LUT construction stage}~\cite{ANNA}, as shown in the bottom middle of \Fig{fig-Example}.
\circled{D} Finally, the query iterates over all the encoded search points that belong to the $nprobs$ chosen clusters to calculate the overall distance. For a given encoded search point with $s$\textsuperscript{th} subspace encoded with codebook entry $e$, assuming it belongs to the $np$\textsuperscript{th} cluster, the total distance is calculated by summing up all the \emph{L2-LUT}$[np][s][e]$ values.
We refer to this third stage as \emph{distance calculation stage}~\cite{ANNA}, as shown in the bottom right of \Fig{fig-Example}.


After calculating the total distance between all encoded search points, the query then sorts the results and selects the top-k closest neighbors. It's important to note that an ANN search may yield false positives and false negatives. For example, the $2^{nd}$ point could be the second closest to the query, but is ignored in ANN search, as depicted in \Fig{fig-Example}.


\begin{figure}[t]
    \centering
      \includegraphics[width=0.99\linewidth]{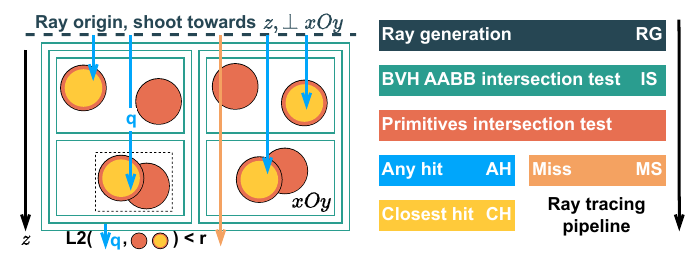}
    \caption{The ray tracing pipeline for RT core.}
    \label{fig-RTPipe} 

\end{figure}

\subsection{Accelerating NN Search with Ray Tracing Core}

Ray tracing (RT) is a distinct rendering pipeline from traditional rasterization~\cite{Gems,Gems2}.
In the RT pipeline, rays are cast from a camera through each pixel into a virtual scene, simulating their interactions with objects to render pixels with accurate colors.
However, the RT pipeline can be computationally expensive and time-consuming due to the need to track interactions between rays and objects, resulting in a large-scale operation.
To accelerate this process, NVIDIA introduced GPUs with dedicated hardware RT accelerators, known as RT cores~\cite{Turing}.
The RT core utilizes a BVH tree-based RT algorithm in hardware, which efficiently finds intersections between lines and surfaces in 3D Euclidean space.
\Fig{fig-RTPipe} illustrates the typical BVH tree-based RT pipeline. 

\revision{R-B}{The NVIDIA Ray Tracing (RT) cores consist of two core hardware components for accelerating the typical ray tracing pipeline: the Axis Aligned Bounding Box (AABB) intersection test and the Bounding Volume Hierarchy (BVH) traversal. 
The AABB-based intersection test first creates bounding boxes (with edges parallel to $x$-, $y$-, or $z$-axes) to bound objects that need to test the intersection with the ray. Then, the ray will conduct a cheap interval-based calculation to test the intersection status with bounding boxes. If a box is intersected with the ray, the ray will further test the intersection status with objects bounded by this tested box. Otherwise, all objects bounded by this box will be ignored. 
Notice that one AABB can recursively contain smaller AABBs, and finally form a tree-like structure with log-scale depth to total objects number, where a node represents an AABB and its child nodes represent smaller AABBs contained by it. This structure is called the BVH tree and makes the process of finding intersected bounding box recursive. Obviously, there can be a huge amount of conditions and divergences in the tree traversal process. The NVIDIA RT core also implements corresponding hardware to accelerate the traversal process of the BVH tree.}

Researchers have utilized the capability of identifying intersections in three-dimensional Euclidean space to apply the RT core in two-dimensional/three-dimensional nearest neighbor search~\cite{RTNN}. 
The basic idea is to first organize $N$ search points into $N$ circles located in the $xOy$ plane, each with a bounding radius $r$. 
Subsequently, the queries can be converted into rays that originate from the query coordinates and are directed towards the z-axis, as illustrated in \Fig{fig-RTPipe}. 
Any circles intersected by a ray indicate that the distance between the query (represented by the ray) and the search points (represented by the circles) is lower than $r$, and thus has the potential to be the nearest neighbors of the query. For example, in \Fig{fig-RTPipe}, the ray $q$ intersects with the two circles in the bottom left quadrant, signifying their potential as the nearest neighbors of the query.


The aforementioned approach, which involves utilizing RT core hardware for accelerating NN search, is limited to low-dimensional (2 and 3) spaces, thereby restricting its practicality. Instead, our work endeavors to explore the effective utilization of RT core for more general NN search tasks, with a specific focus on ANN search in high-dimensional spaces.


\section{Motivation}
\label{sec:Motivation}

This section provides an analysis of the state-of-the-art library of high-dimensional approximate nearest neighbor (ANN) search, FAISS~\cite{FAISS}. We begin by measuring the breakdown of execution time for FAISS queries, followed by an analysis of the identified inefficiencies. Finally, we propose optimization takeaways based on the findings of our analysis.
 


\subsection{Execution Time Breakdown}

We use the latest version of FAISS~\cite{FAISS} and the DEEP1M dataset~\cite{DEEP1M} in our study. 
Specifically, we configure FAISS with \texttt{<IVF4096,PQ48>}, where 1,000,000 search points are grouped into 4096 clusters, and the 96-dimensional space is divided into 48 2-dimensional subspaces. 
We measure the execution time of three stages as mentioned in \Sec{sec:IVFPQ} (\emph{filtering}, \emph{L2-LUT construction}, \emph{distance calculation}) on an NVIDIA Geforce RTX 4090 GPU~\cite{RTX4090}.
For each query, we evaluate its execution time with different $nprobs$, which is a hyper-parameter and means the number of selected clusters in \emph{filtering}.
 

\begin{figure}[t]
  \centering
    \includegraphics[width=0.49\linewidth]{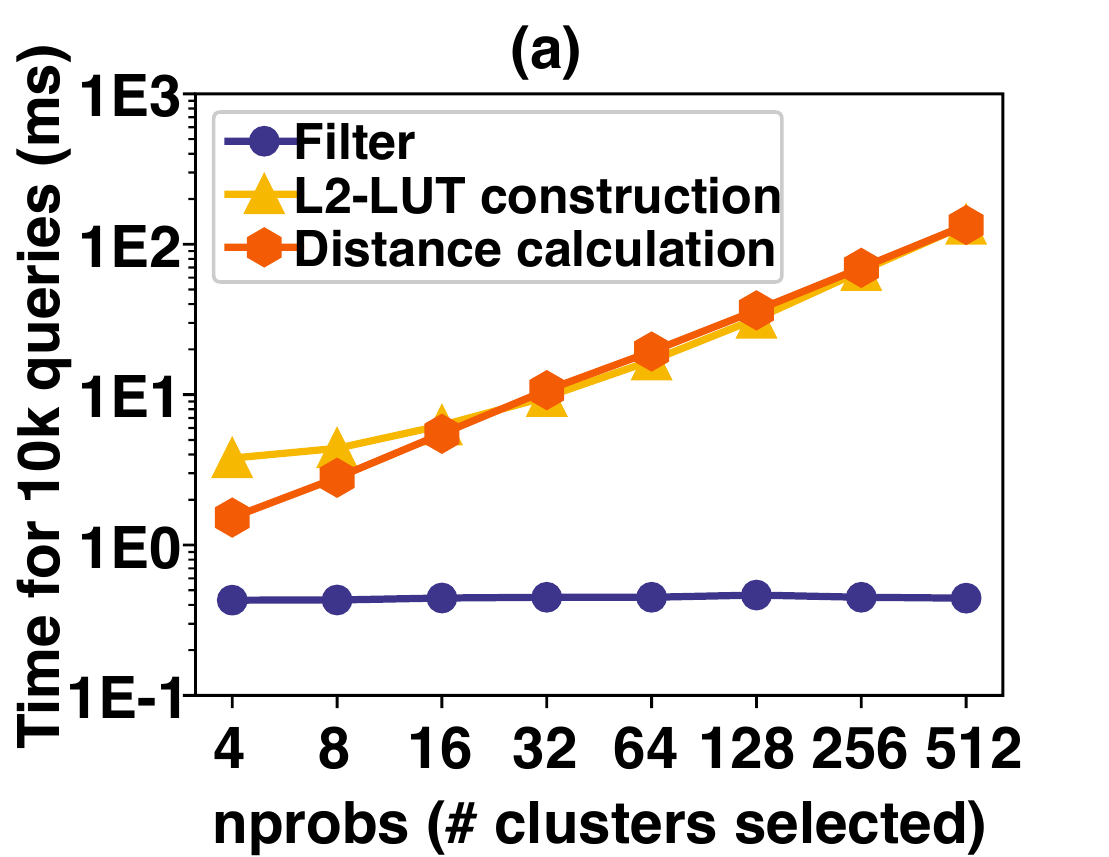}
    \includegraphics[width=0.49\linewidth]{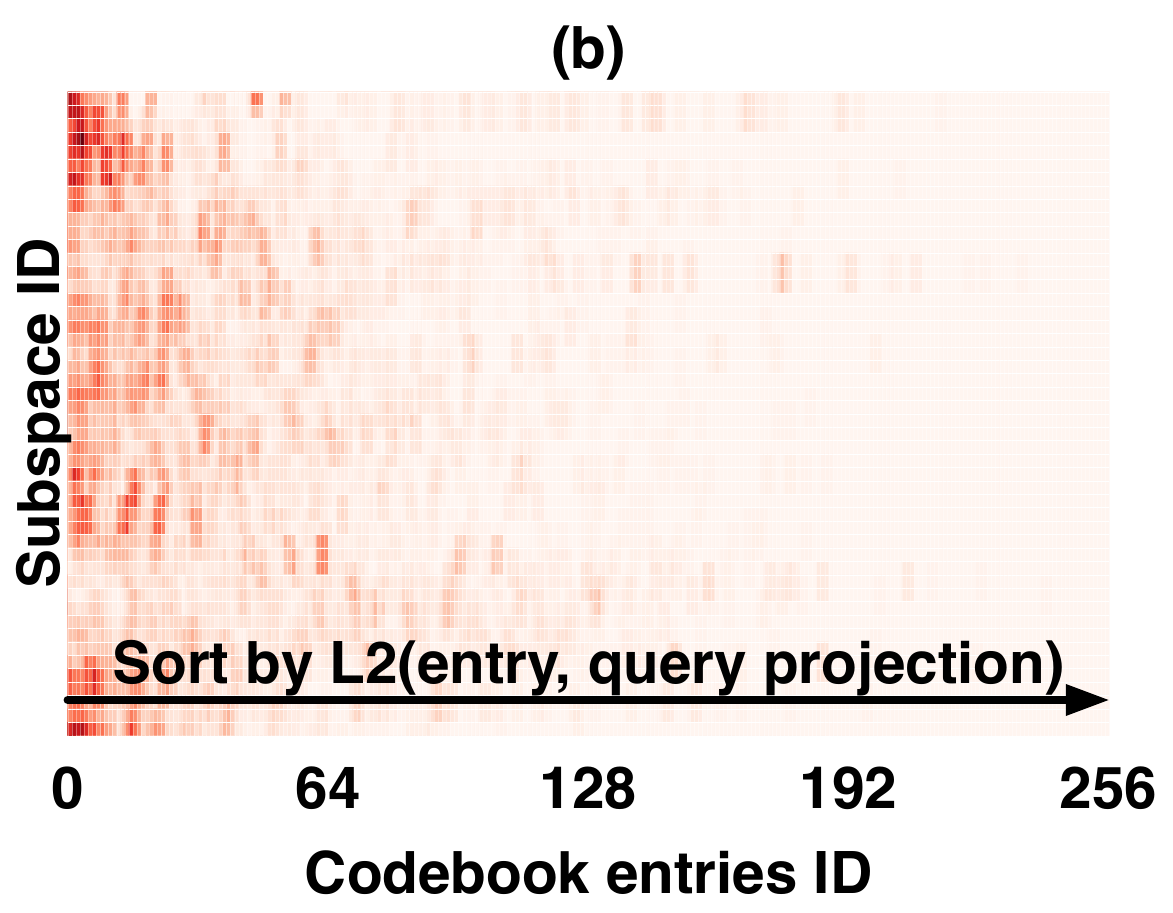}
  \caption{(a) Execution time breakdown of searching queries in DEEP1M using FAISS. (b) Codebook entries usage of a single query: higher usage frequency leads to darker color.}
  \label{fig-Breakdown} 

\end{figure}

The experimental results are presented in Figure \ref{fig-Breakdown}(a), which illustrates the breakdown of execution time. The majority of the execution time is consumed by the \emph{L2-LUT construction} and \emph{distance calculation} stages, accounting for approximately 90\% to 99.9\% of the total time. Additionally, the time taken by these stages increases linearly with the hyper-parameter $nprobs$, which is set smaller for higher performance and larger for better search quality~\cite{EarlyQuit}. 
On the other hand, the \emph{filtering} stage remains relatively stable, as its computation depends on $Q \times D \times C $, where $C$ is independent of $nprobs$. This observation motivates us to focus on optimizing the \emph{L2-LUT construction} and \emph{distance calculation} stages, as we will analyze the inefficiencies in the current approach in below.

\begin{figure}[t]
  \centering
    \includegraphics[width=0.49\linewidth]{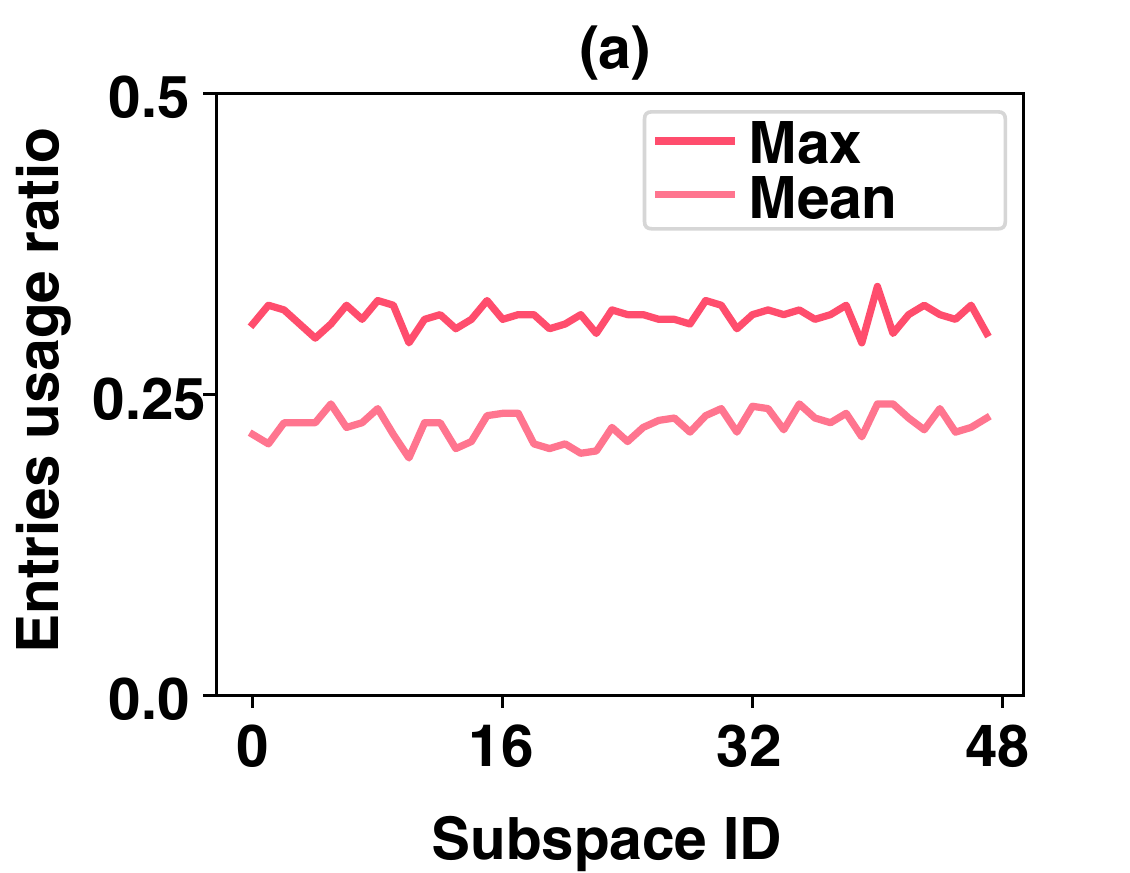}
    \includegraphics[width=0.49\linewidth]{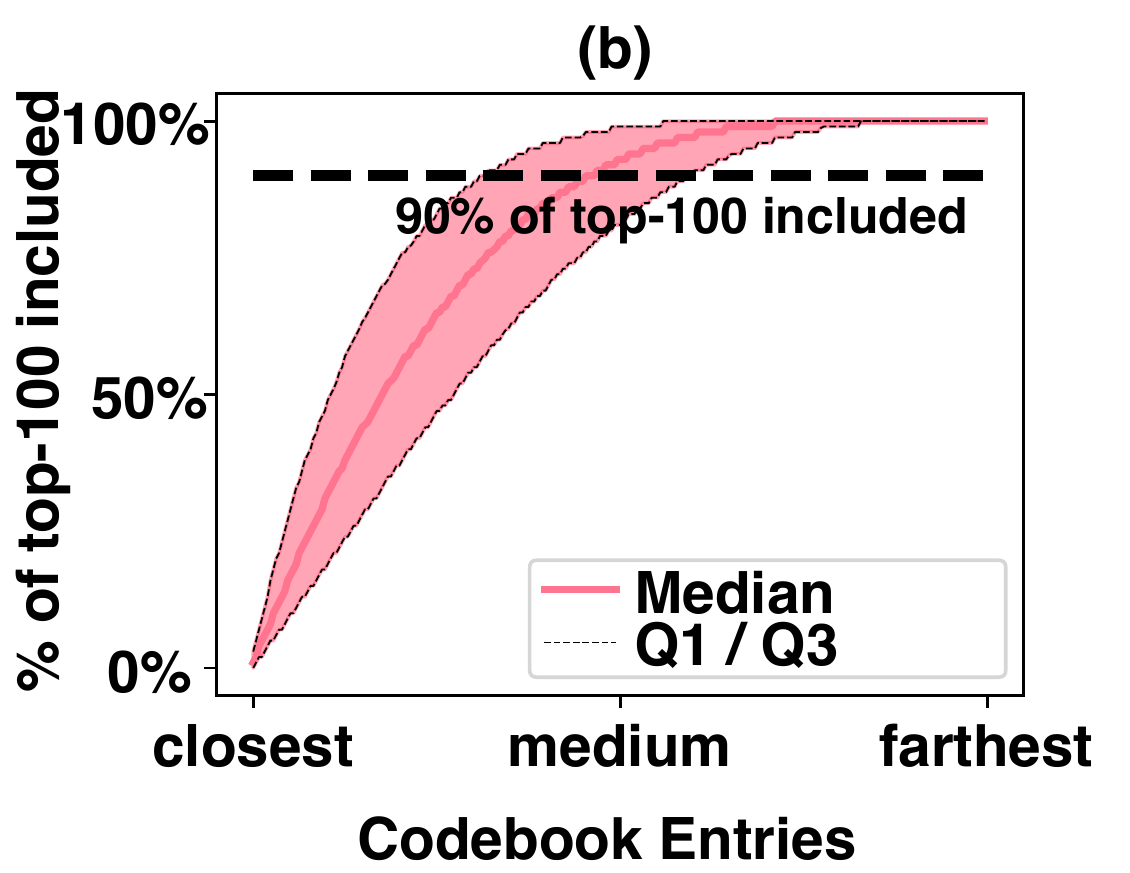}
  \caption{(a) Max used ratio of codebook entries on every sub-dimension of 100 queries. (b) CDF of entries to contain top-100 from closest to farthest. Figures are ploted using DEEP1M dataset.\\ \emph{*Q1, Q3 are $1^{st}$ (25\%), $3^{rd}$ (75\%) quantile (percentile).}}
  \label{fig-Motivation} 

\end{figure}

\subsection{Sparsity of Codebook Entries Used by Top-k Neighbors}
\label{sec:Motivation_2}

The current ANN search implementation, like FAISS, calculates the pairwise distance between the query projection and codebook entries in all subspaces during the \emph{L2-LUT construction} stage. However, our findings indicate that for a single query, only limited codebook entries are used to identify the top-100 search points with the closest proximity.

To illustrate the above point, we calculate the usage frequency of each codebook entry by the top-100 search points. 
The results in the form of heatmap are presented in \Fig{fig-Breakdown}(b), where the statistics for all entries from different subspaces are depicted. 
The shading of the cells corresponds to the frequency of usage, ranging from 0 to 100, with darker colors indicating higher frequency of usage. 
For example, a $cell\lbrack 31\rbrack\lbrack 114\rbrack=75$ means that in the $31^{st}$ subspace, 75 among top-100 search points are encoded using the $114^{th}$ codebook entry.
A value of 0 signifies that none of the top-100 search points are encoded with the respective codebook entry.

\begin{figure}[H]
  \centering
 \includegraphics[width=0.489\linewidth]{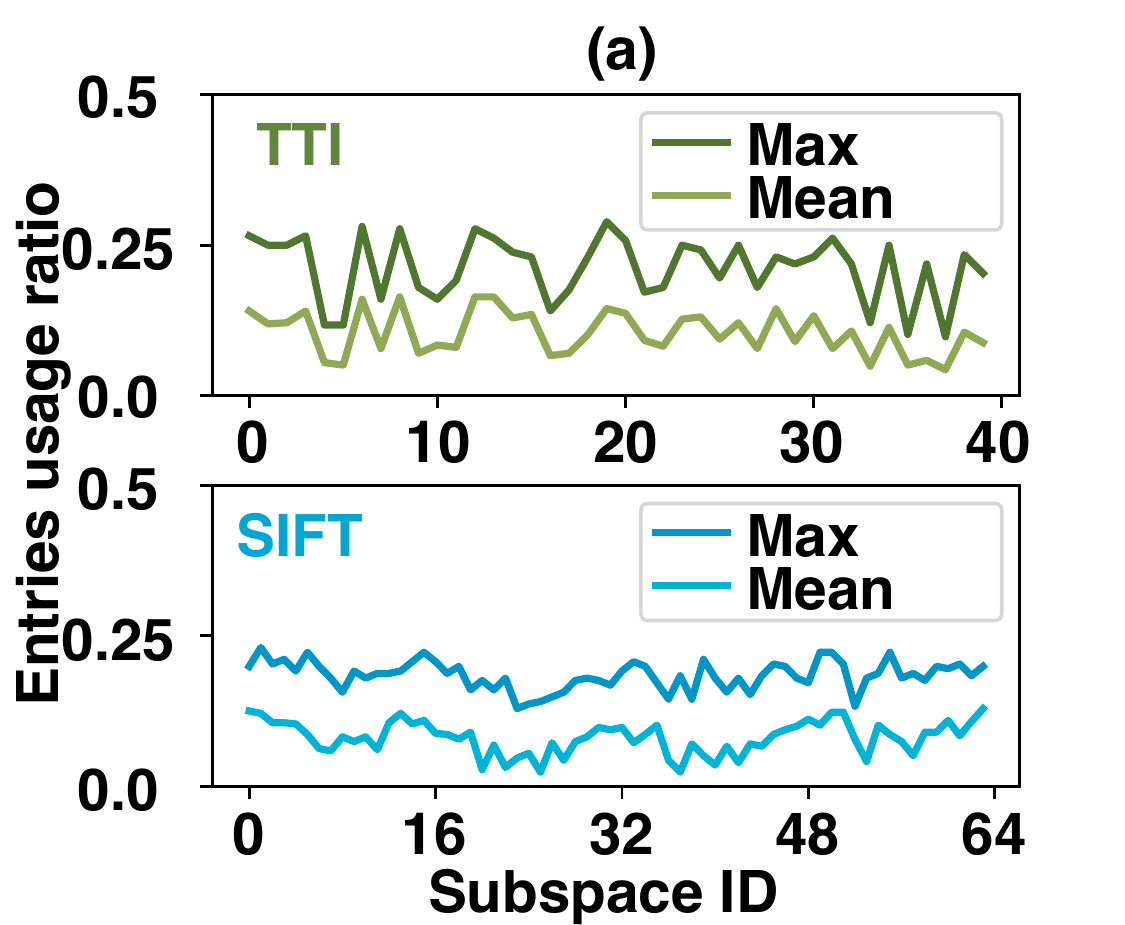}
 \includegraphics[width=0.491\linewidth]{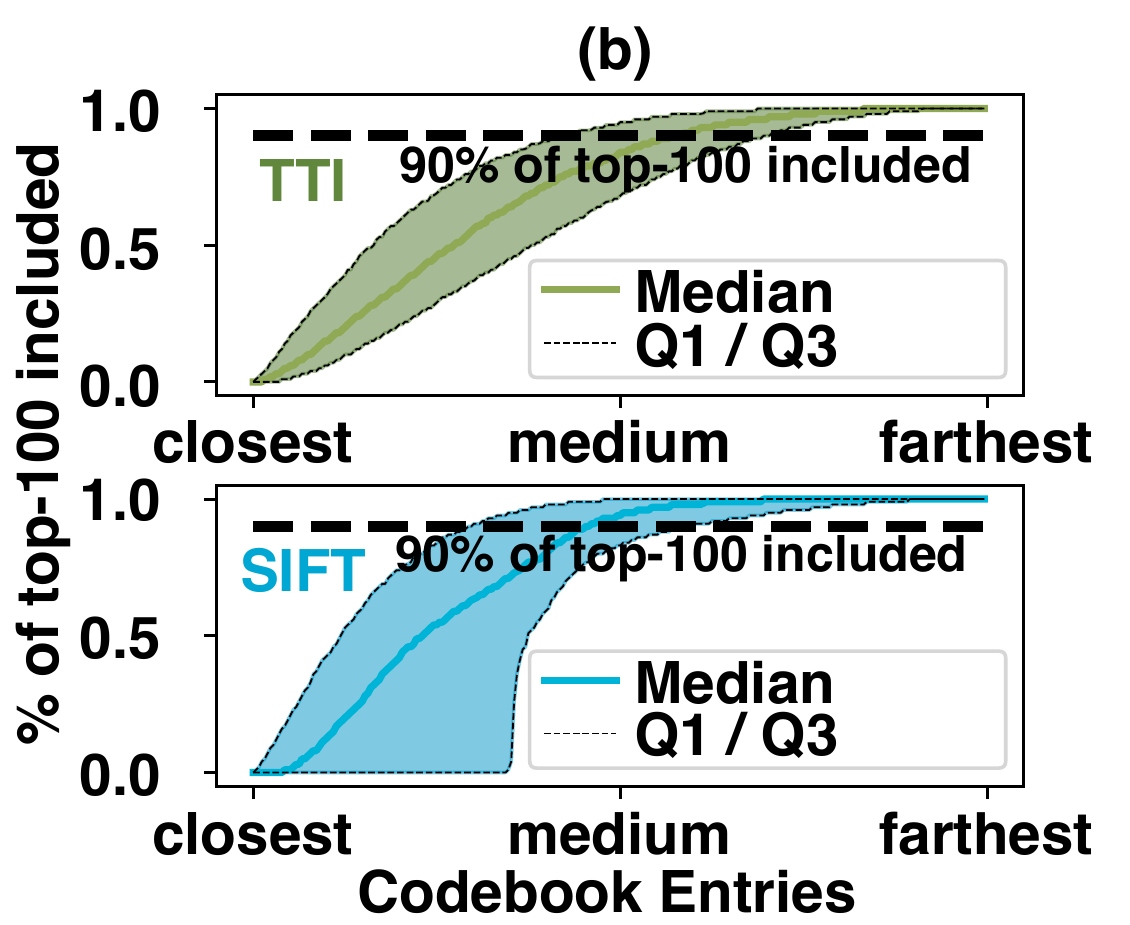}
    \caption{(a) Codebook entries average and maximal usage ratio and (b) CDFs of entries to contain top-100 from closest to farthest.}
    \label{fig-DSSparsityCDF} 

\end{figure}

The codebook entries utilization in each subspace for the DEEP1M dataset~\cite{DEEP1M} with the \texttt{PQ48} configuration is shown in \Fig{fig-Motivation}(a), indicating that, on average, only 25\% of the entries (at most 30\%) are utilized. 
To investigate the impact of data distribution on this sparsity, we analyze the SIFT1M (\texttt{PQ64})~\cite{SIFT1M} and TTI1M~\cite{SIFT1M} (\texttt{PQ40}) datasets, as presented in \Fig{fig-DSSparsityCDF}(a). 
It can be observed that, on average, these datasets also exhibit less than 30\% codebook entries utilization. 
Exploiting this sparsity can potentially result in a significant reduction of up to 70\% floating point operations in pairwise distance calculation for each query, thereby substantially reducing the time required for the \emph{L2-LUT construction} stage.


With above analyses, we derive the first key takeaway:
\textbf{Codebook entries used by top-100 neighbours are sparse.}


\subsection{Spatial Locality of Codebook Entries Used by Top-k Neighbors}

Although we have shown that there exists a high sparsity degree in the codebook entries used by top-k points, it maybe still challenging to convert the sparsity to actual performance speedup as sparsity often results in irregular access patterns.
However, as depicted in \Fig{fig-Breakdown}(b), the codebook entries that are actually used are densely concentrated in the front half of the frequency heatmap. This suggests that the used codebook entries are closer to the query projection compared to others, as the heatmap is sorted based on the distance between the entry and the query projection in each subspace.


To validate this claim, we calculate and plot the cumulative distribution function (CDF) of actual top-100 search points from the closest to farthest entries in every subspace. 
As depicted in \Fig{fig-Motivation}(b), we observe that considering using half of the codebook entries enables us to obtain over 90\% of the top-100 search points. 
The CDFs for the SIFT1M and TTI1M datasets are shown in \Fig{fig-DSSparsityCDF}(b), respectively. Despite having different patterns, both datasets exhibit a similar property, with approximately 50\% of the closest entries containing over 90\% of the top-100 ground truth.


With above analyses, we derive another key takeaway:  
\textbf{Codebook entries used by top-100 neighbours are closely distributed in the space.}
\section{Selective L2-LUT Construction on RT Core}
\label{sec:AlgoMap}

In this section, we present our algorithm enhancement and RT core mapping for an efficient alternative to the original \emph{L2-LUT construction} stage.
We begin by introducing our algorithmic intuition and design details, then we introduce how we map the enhanced algorithm to the RT core.

\subsection{Threshold Based Selective L2-LUT Construction}
\label{sec:ThresholdBasedL2LUT}

In this subsection, we describe and explain our algorithm design to leverage the features above of the current ANN search. The new algorithm aims to replace current \emph{L2-LUT construction} in our searching pipeline and provide significant operation saving with acceptable approximation.

\paragraph{Intuition of Algorithm Enhancement.}
\label{sec:Intuition}
Our algorithm aims to exploit the sparsity and spatial locality in subspaces, resulting in improved search performance through additional approximation techniques alongside product quantization. 
Specifically, we propose a distance checking-based approach to efficiently prune unnecessary entries in each subspace. This approach involves setting distance thresholds in all subspaces, defining the interested region of a query projection as the union of all points with distances from the query projection being less than the distance threshold, and subsequently discarding codebook entries that fall outside the interested region of a query projection. This pruning strategy is motivated by the findings presented in \Sec{sec:Motivation}.

\begin{wrapfigure}{r}{4.5cm}
    \centering
    \includegraphics[width=0.99\linewidth]{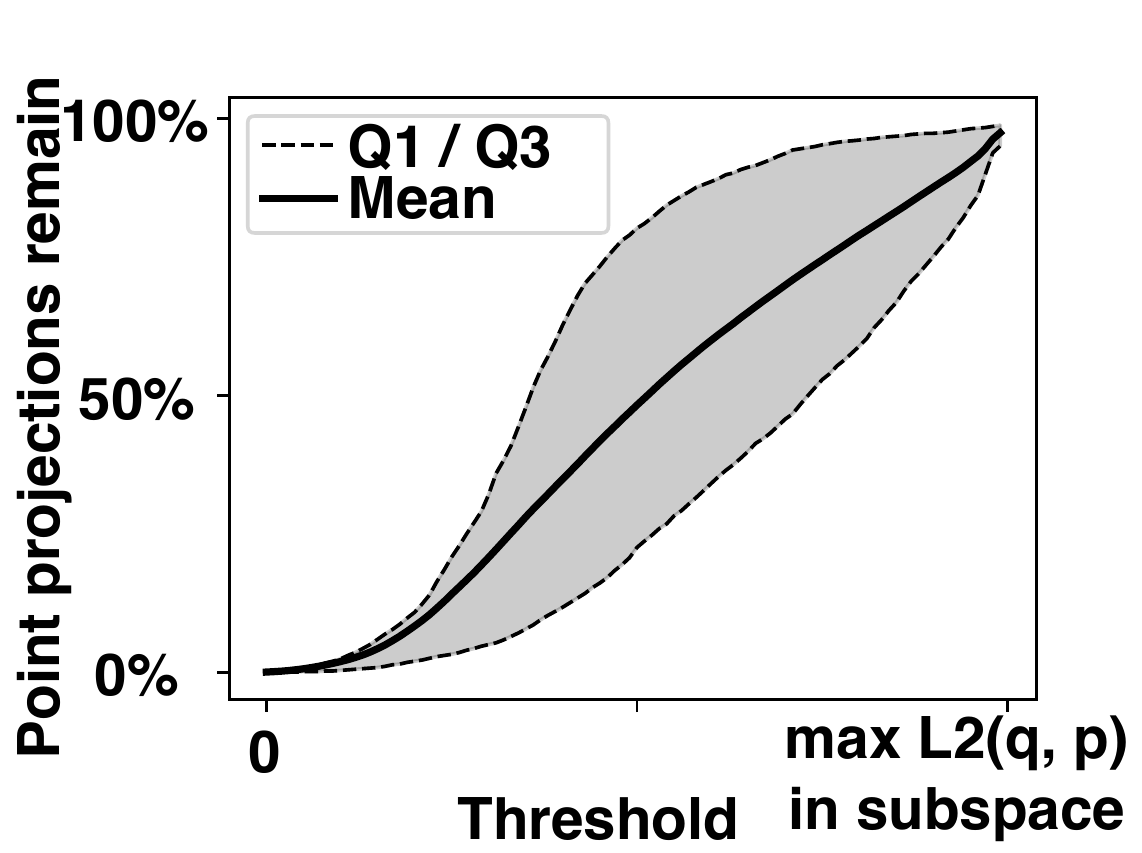}
    \caption{Remained search points that need to accumulate the distance.}
    \label{fig-SavingOp} 
\end{wrapfigure}

Pruning unnecessary codebook entries can significantly reduce the number of distance calculations. 
As in \Fig{fig-SavingOp}, the remaining search point projections that require accessing the L2-LUT and distance calculations decrease linearly with the threshold. 
This finding suggests that the projections of the top-100 search points are closely distributed around the query projection in a subspace. 
Consequently, a substantial amount of L2-LUT lookups and distance calculations can be saved by filtering out codebook entries far away from the query projection in each subspace.

\paragraph{Choose Necessary Entries with Efficiency.}
As discussed in \Sec{sec:Intuition}, our algorithm leverages the sparsity and spatial locality by selectively choosing the codebook entries that fall within the region of interest in each subspace. 
The distances between these selected entries and the query projection are then calculated for constructing a \emph{L2-LUT} in a selective manner. To achieve this, we bound smaller entries/boxes into a larger box and hierarchically organize these bounding boxes into a tree structure. As such, we can avoid pairwise distance calculations between all $E$ entries and the query projection in each subspace, and instead perform inside/outside checks with logarithmic complexity of $logE$.

\paragraph{Determining Proper Threshold.}
The selection of an appropriate threshold is crucial in our approach, as it is used to distinguish whether an entry falls within the interested region of a query projection. A tight threshold may filter out too many entries, resulting in missed true neighbors, while a relaxed threshold may include unnecessary entries, leading to a waste of time and resources. To determine the optimal threshold, we conduct a thorough study of the relationship between query features and the threshold that can contain the top-100 search points in each subspace.

\begin{figure}[b]

    \centering
      \includegraphics[width=0.485\linewidth]{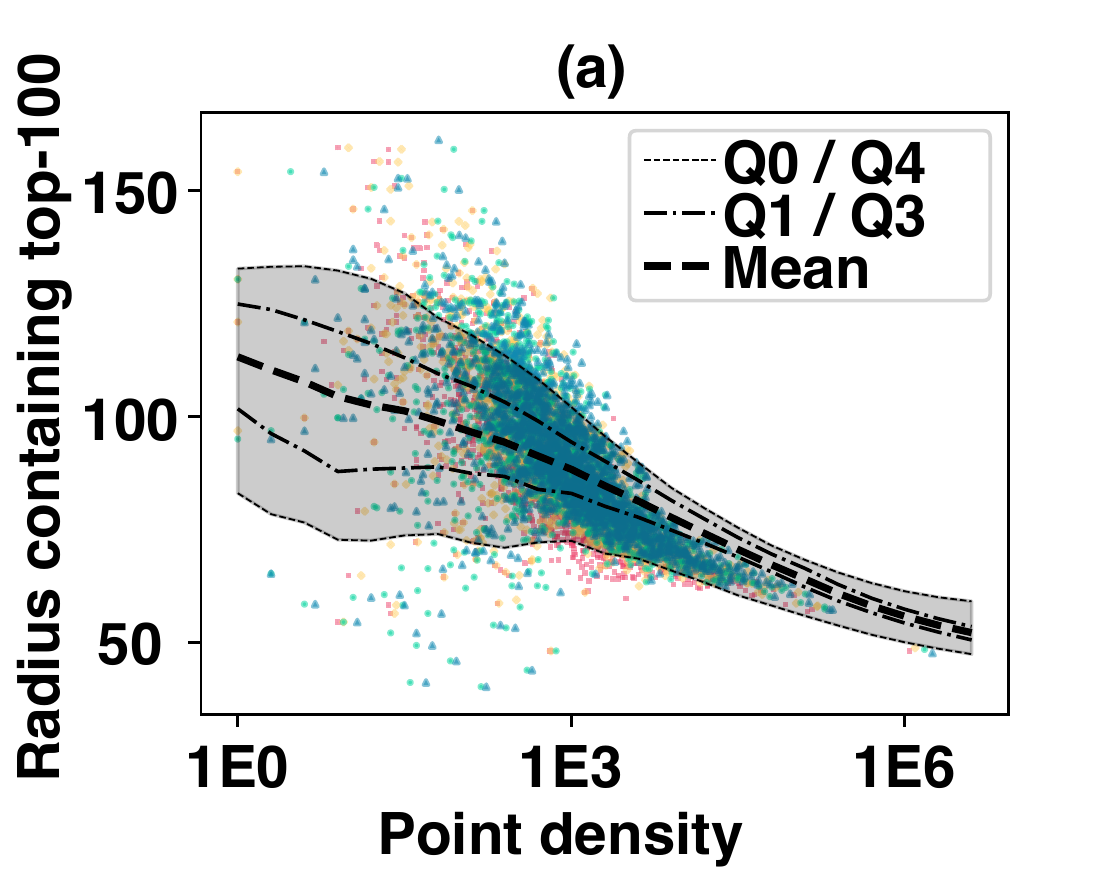}
      \includegraphics[width=0.49\linewidth]{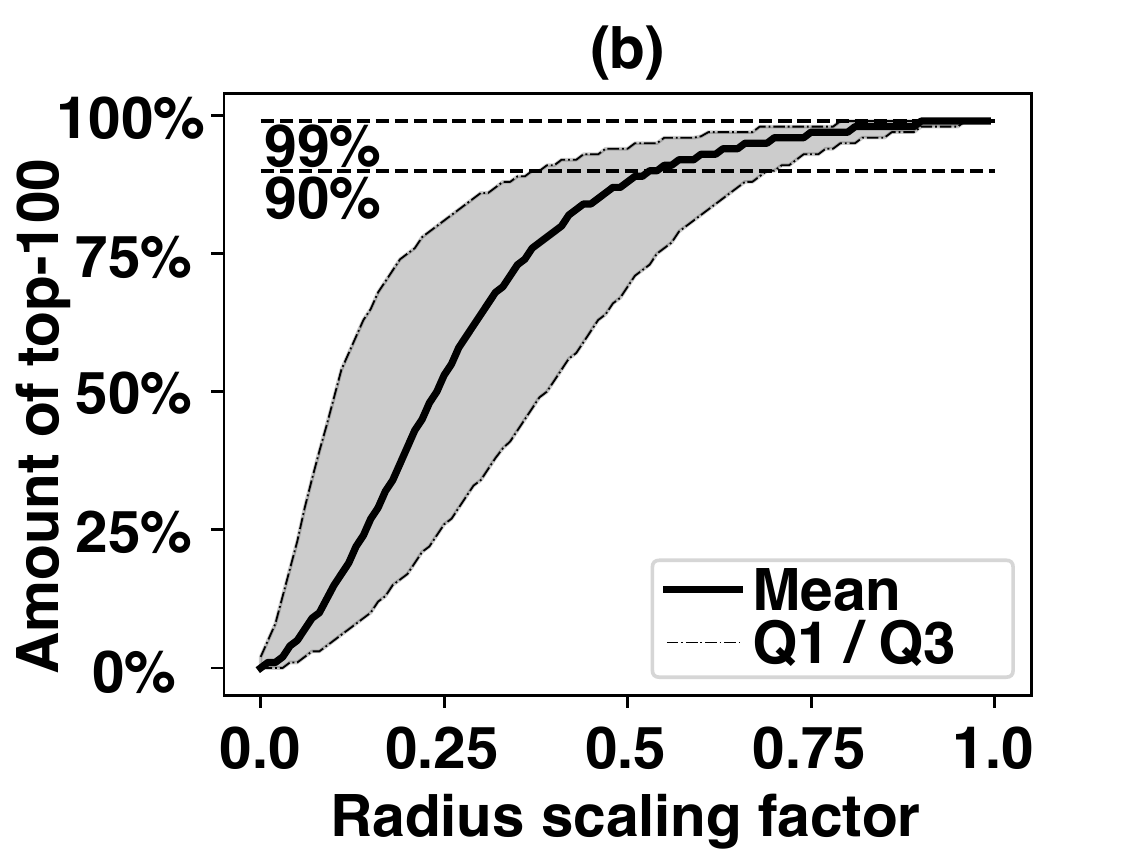}
    \caption{(a) The relation between threshold to contain top-100 search points and region density a query falls into. (b) The amount of top-100 search points contained when threshold scales smaller.\\ \emph{*Q0$=$Q1$-1.5\times$IQR, Q4$=$Q3$+1.5\times$IQR, IQR$=$Q3$-$Q1.}}
    \label{fig-radius} 
    
\end{figure}

As shown in \Fig{fig-radius}(a), we observe a negative correlation between the threshold for containing the top-100 search points and the region density of the query projection in a subspace. To calculate the region density, we divide the entire subspace into a $100 \times 100$ grid and define the density of each cell as the quotient obtained by dividing the number of search point projections falling into that cell by the area of the cell. This finding is reasonable, as a higher threshold is likely to contain many search point projections in high-density regions, resulting in a higher probability of containing the top-100 search points. Conversely, in low-density regions, even the same threshold may not be able to contain 100 search point projections, let alone the top-100 points.

Based on our findings, we propose incorporating dynamic radius mechanics to determine an appropriate threshold during runtime. Initially, we generate a density map and a simple regression model offline. The density map consists of a $100\times 100$ grid for each subspace, where each cell records the density computed as previously described. Subsequently, we randomly select several search point projections to train the regression model, with the region density as input and the threshold to contain the top-100 search points as output. During runtime, we look up the density of the query projection in each subspace and use the regression model to infer a query-specific threshold. Notably, we find that a simple polynomial regression model accurately captures the relationship, resulting in minimal runtime overhead for threshold determination. Once the threshold is determined, we disregard all codebook entries beyond the threshold and compute the distance between the query projection and the remaining codebook entries to create an L2 lookup table (L2-LUT).

Furthermore, we have observed a power-law pattern in the variation of the threshold value, as illustrated in \Fig{fig-radius}(b). When the distance threshold is scaled down to half of its original value to accommodate the top-100 search points, approximately 90\% of these points are retained. This indicates the potential for using a smaller distance threshold to further prune codebook entries. In other words, we can trade-off slightly lower search quality for significantly enhanced search throughput, such as queries per second (QPS). In our work, we provide users with the flexibility to set the threshold value through a dedicated interface, enabling them to make this trade-off according to their specific needs.

\subsection{Mapping the Selective Algorithm to the RT Core}

In this subsection, we explain how to map the computation of selective L2-LUT construction described in \Sec{sec:ThresholdBasedL2LUT} to the ray-tracing (RT) core in modern GPUs.
The key approach is to utilize the RT core to accurately, efficiently determine the intersections of rays with the selected search points within each subspace. 
Additionally, we exploit the capabilities of the RT core for efficient hit distance calculation, which can be extended to support other distance metrics like inner product.



\paragraph{Algorithm Mapping.}
The RT core implements bounding box intersection check (AABB) and hierarchical data structure traversing (BVH) in the hardware, which aligns with the computational characteristics of our proposed threshold-based LUT construction.
Similar to the original ANN search algorithm discussed in \Sec{subsec:ann_background}, our RT-core-based ANN algorithm comprises both an offline and online component.

\begin{figure}[t]
    \centering
      \includegraphics[width=0.99\linewidth]{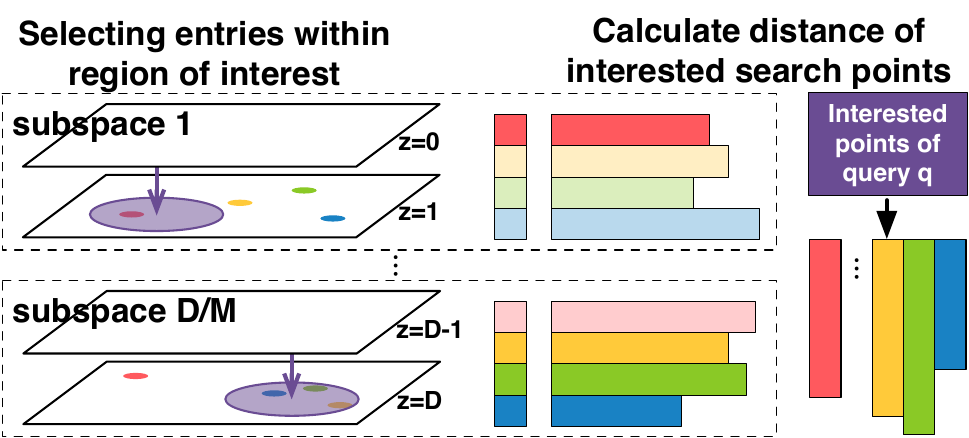}
    \caption{Intuition of hardware mapping of \proj{}.}
    \label{fig-Intuition} 
    
\end{figure}

Previous research~\cite{RTNN} has demonstrated the efficient inside/outside check capability of the RT core through its BVH tree-based algorithm. With this capability, our method has the potential to save computational resources in determining whether an entry falls inside the region of interest within a query projection in a subspace. To implement this, we align our distance-checking approach with the RT core by placing a sphere at the coordinates of the query projection, with a radius set to the distance threshold. This sphere naturally defines the region of interest. Subsequently, rays are cast from codebook entries towards the region of interest in the RT core, enabling rapid determination of whether an entry falls within the region of interest, as the left part of \Fig{fig-Intuition} shows.


In our approach, we utilize a sphere to delineate the region of interest in a query projection. However, this approach necessitates adding spheres into the runtime scene, resulting in excessive overhead. To mitigate this issue, we take advantage of the commutativity of the L2 distance metric. 
Specifically, we pre-generate spheres at the center of each codebook entry during the  offline processing stage, and cast query projection towards these spheres at runtime.
This yields identical inside/outside results as the online sphere construction.


Next, we construct the L2-LUT specifically for the entries located within the region of interest. 
In this process, we also utilize RT cores for the fast and efficient distance calculation of these selected entries, employing the concept of hit distance which will be detailed in the next paragraph. Once the L2-LUT is constructed, it can be employed for performing distance lookup and accumulation for the points of interest in the search, as demonstrated in the right part of \Fig{fig-Intuition}.



\paragraph{Calculating Hit Distance.}
The construction of L2-LUT requires calculating the distance between each hit entry and the sphere center. The naive approach is to directly use the coordinates of the hit entries stored in global memory, which may have irregular memory accesses due to the sparsity of hit entries. To address this issue, we propose utilizing the results from the RT core to efficiently calculate the distance, thus avoiding global memory accesses.


The RT core incorporates the notion of time in ray tracing, signifying the duration of a ray's travel. By default, a ray traverses one unit of space within a one unit time interval. 
There are two defined crucial time points: $t_{hit}$ and $t_{max}$. The former denotes the time interval from the initiation of ray travel to the point when it intersects an object, while the latter represents the maximum duration that a ray can travel. We can obtain $t_{hit}$ in hit shader without global memory access. 
 
\begin{figure}[t]
  \centering
    \includegraphics[width=0.99\linewidth]{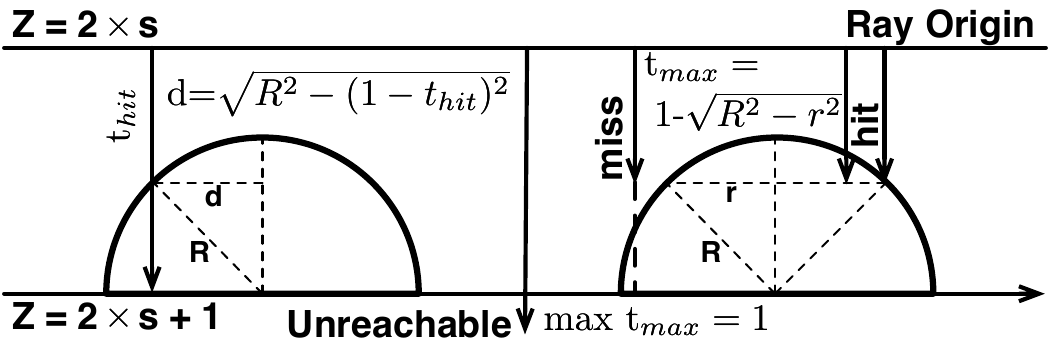}
    \caption{(left) We use the $t_{hit}$ to calculate the hit distance between query projection and sphere centroid. (right) We dynamically adjust the radius of the sphere that a ray can hit by adjusting $t_{max}$. }
  \label{fig-CalcDistDynR} 
\end{figure}

To accurately calculate the distance, we employ the $t_{hit}$ of a ray and rapidly determine the distance between the hit point and the center of the sphere using the radius of the hit sphere. This allows us to obtain the precise distance between the query projection and the codebook entry, as depicted in the left portion of \Fig{fig-CalcDistDynR}.
Notice that $R$ in the figure means the radius of the spheres and are now set as identical constants. 




\paragraph{Dynamic Threshold.}
As mentioned in the previous subsection, our approach also incorporates a dynamic distance threshold to obtain top-100, which can also be adjusted by the user to balance search quality and throughput. A naive implementation would involve modifying the radius of spheres at runtime, which may result in unacceptable overhead.


To eliminate the need for online sphere creation, we convert the dynamic distance threshold into a dynamic maximum travel time for rays. For smaller regions, we set a correspondingly smaller value for $t_{max}$, as shown on the right side of \Fig{fig-CalcDistDynR}. For example, if the original threshold is $0.6$, a value of $t_{max}=0.64$ corresponds to a scaling factor of $0.8$. 
Thus, we support both dynamic distance threshold and user-defined scaling factor by adjusting $t_{max}$ only, and we can set the radius of all spheres to be identical, which will significantly simplify the aforementioned distance calculation and further processing.


\paragraph{Inner Product Similarity Support.}

Our approach efficiently supports maximal inner product similarity (MIPS) at minimal computational cost through a well-designed transformation mechanism. 
Previous methods rely on introducing extra dimensions so that they can maximize inner product by minimizing L2 distance between transformed query and search points~\cite{L2IP1,L2IP2}. 
The extra dimensions affect training, aligning, search performance~\cite{FAISS}. 
Instead, we propose a method that is free of extra dimensions and RT-core friendly. 



Notice that we use $t_{hit}$ to calculate $L2(e,q)$ in a 2D subspace where $e$ is the codebook entry, that is:
\vspace{-0.2cm}
\revision{}{
\begin{equation*}
\label{eq:mips3}
\begin{aligned}
L2^{2}(\bm{e},\bm{q})&=R^{2}-(1-t_{hit})^{2}=(x_{e}-x_{q})^2+(y_{e}-y_{q})^2\\
IP(\bm{e},\bm{q})&=x_{e}x_{q} + y_{e}y_{q}=(\underline{x_{e}^2}+\underline{y_{e}^2}+x_{q}^2+y_{q}^2-L2^{2}(\bm{e},\bm{q}))/2\\
&=(\underline{x_{e}^2}+\underline{y_{e}^2}+x_{q}^2+y_{q}^2-R^2 + (1-t_{hit})^2)/2
\end{aligned}
\end{equation*}}
The codebook entry related part $x_{e}^2,y_{e}^2$ requires accessing global memory lookup for their coordinates at runtime to calculate inner product $IP(e,q)$. 
While with the RT core, we can eliminate $x_{e}^2,y_{e}^2$ by replacing $R$ with $R^{'}=\sqrt{R^2+x_{e}^2+y_{e}^2}$ without any extra dimension(s) as follows:
\vspace{-0.2cm}
\revision{}{
    \begin{equation*}
        \begin{aligned}
            t_{hit}^{new}&=1-\sqrt{R'^2-\underline{x_{e}^2}-\underline{x_{e}^2}-x_{q}^2-y_{q}^2+2\times IP(\bm{e},\bm{q})}\\
            &=1-\sqrt{R^2-x_{q}^2-y_{q}^2+2\times IP(\bm{e},\bm{q})}\\
            &\Rightarrow IP(\bm{e},\bm{q})=(x_{q}^2+y_{q}^2-R^2+(1-t_{hit}^{new})^2)/2
        \end{aligned}
    \end{equation*}
}
\revision{R-S}{where $x_q, y_q, R$ are all constant values for a single query.}
Thus, the inner product $IP(e,q)$ can be directly calculated using $t_{hit}^{new}$, without any sphere coordinate accesses.
Furthermore, the term $x_{q}^2+y_{q}^2$ can be disregarded as it remains constant for all codebook entries. Consequently, only the radiuses of spheres need to be adjusted from $R$ for the L2 distance metric to $\sqrt{R^2+x_{e}^2+y_{e}^2}$ for the inner product metric offline.
Notice that we also need to change metric of the cluster in \emph{filtering} from L2 distance to the inner product.






\begin{figure}[t]
    \centering
      \includegraphics[width=0.99\linewidth]{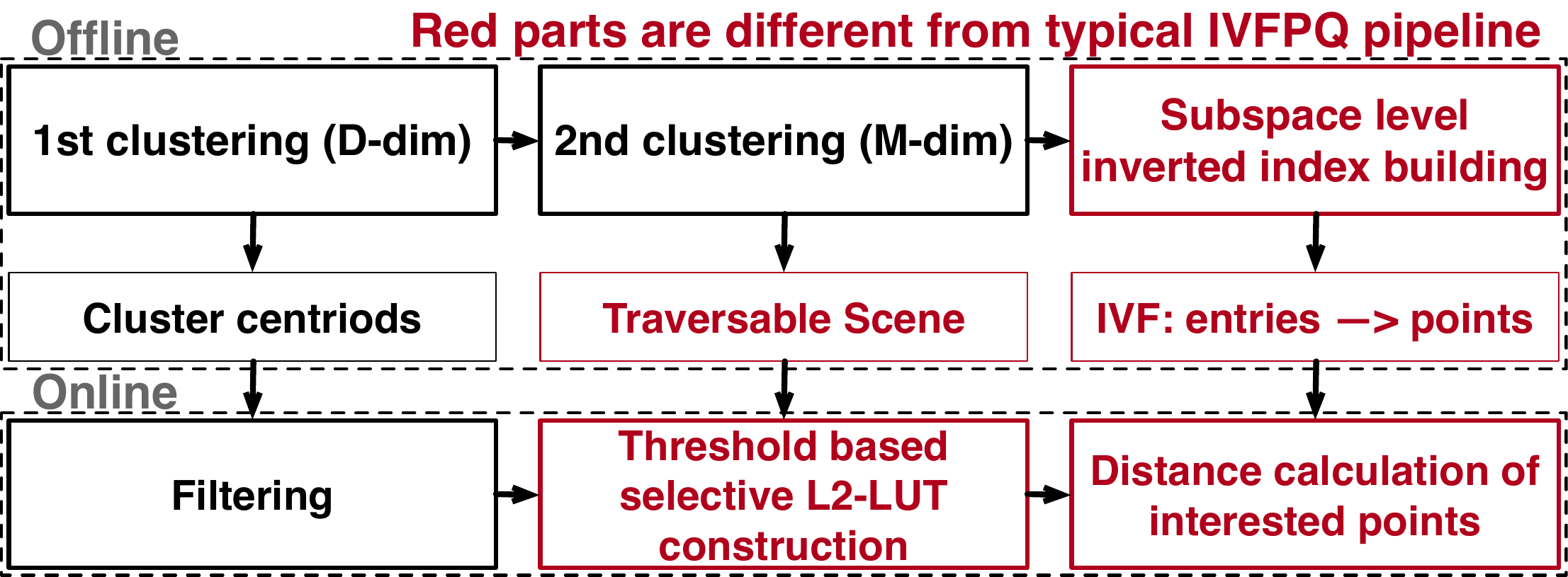}
    \caption{Overview design of \proj{}.}
    \label{fig-Overview} 
\end{figure}

\section{\proj{} System Design}

Based on aforementioned insights, we propose \proj{}, an end-to-end search system for efficient ANN search in high-dimensional space. 
\proj{} consists of offline and online phases leveraging the algorithmic enhancement and hardware mapping mentioned in \Sec{sec:AlgoMap}, which are described in \Sec{sec:OfflinePhase} and \Sec{sec:OnlinePhase}. Besides, we propose pipelining and aggressive approximation leveraging the hardware features of the RT core, which are discussed in \Sec{sec:Pipeline} and \Sec{sec:Aggressive}.

\subsection{System Design Overview}

We present the overview of \proj{} in \Fig{fig-Overview}, which is also formally described in supplemental materials by algorithm `\proj{} end to end'.
In the offline phase, we prepare the traversable scene and inverted indices from codebook entries to search points in every subspace. 
Once a batch of queries arrive, we first conduct the filtering that is identical to the original \texttt{IVFPQ} approach. 
Then we use the RT core to do threshold-based selective L2-LUT construction to obtain only necessary entries falling inside the interested region of query projection in every subspace. 
Finally, we conduct the distance calculation where only interested search points are considered for the final accumulated distance, by using the inverted indices prepared offline and L2-LUT constructed online.

\subsection{Offline Preparation Phase}
\label{sec:OfflinePhase}
In the offline phase, we need to prepare a traversable scene and subspace-level inverted indices for later online searching. 
\Alg{Algo-BuildIndex} shows its details.
We first use the typical \texttt{IVFPQ} offline training process.
Specifically, we first obtain the $C$ cluster centroids and labels of search points needed by \emph{filtering} (line 3-4), and then generate the codebook trained by the residual between search points and their centroids (line 5-9). 

The conventional \texttt{IVFPQ} approach encodes and stores all search points with the codebook entries, which is not efficient for our selective L2-LUT construction. To address this issue, we maintain an inverted index from codebook entries to search points (lines 14-16) in each subspace. For example, $Map[114][19][24]$ contains all search points that satisfy the following conditions: i) the search point belongs to the $114^{th}$ cluster, and ii) its projection is encoded with the $24^{th}$ codebook entry in the $19^{th}$ subspace. This inverted index enables us to only iterate through necessary search points from entries that are close to the query projection in a subspace.


In the $d^{th}$ subspace, the spheres representing the entries of this subspace are positioned at the corresponding $x$ and $y$ coordinates. For the $z$ coordinate, we place the entries from different subspaces at different depths, specifically $z=2s+1$ for the $s^{th}$ subspace. This approach prevents interference from rays originating from other subspaces during the ray tracing process. 
Lines 10-13 show the codebook entries placement in $D/M$ subspaces, which is also illustrated in the left of \Fig{fig-Intuition}.


Besides the position of spheres, we also need to determine their radius (i.e., distance threshold). 
Recall that the distance threshold depends on the density of the grid a query projection falls into (\Sec{sec:ThresholdBasedL2LUT}), and we use the travel time-based method to enable dynamic threshold. 
So, we set the same radius for all spheres for the convenience of further computation. 

\begin{algorithm}[b]
    \small
        \caption{Build a traversable scene, prepare cluster centroids of filter and entry-search points mapping offline.}
        \label{Algo-BuildIndex}
        \textbf{Input:}  $points\lbrack N\rbrack \lbrack D \rbrack$, $M=2$, $E$, $C$, $metric$\\
        \textbf{Output:} $Map\lbrack C\rbrack\lbrack \frac{D}{M}\rbrack\lbrace entry\_id : points\_id\lbrack~\rbrack  \rbrace$\\
        \textbf{Output:} $centroids, labels$, $Scene$
        \begin{algorithmic}[1]
            \Function {BuildRTScene}{$points\lbrack N\rbrack \lbrack D \rbrack$, $M$, $E$, $C$}
            \State{$Scene,filter\leftarrow \emptyset, kmeans(points, n\_cluster=C)$}
            \State{$centroids, labels\leftarrow filter.centroids, filter.labels$}
            \State{$residual\leftarrow\lbrack x-centoids\lbrack x.label\rbrack for~x~in~points\rbrack$}
            \For{$s\in\lbrack 0, \frac{D}{M})$}
                \State{$res\leftarrow residual\lbrack:,2s:2s+2\rbrack$}
                \State{$codebook[s]\leftarrow kmeans(res, n\_cluster=E)$}
                \State{$entries\leftarrow codebook[s].centroids$}
                \For{$e\in[0,E)$}
                    \State{$x,y,z\leftarrow entries[e].x,entries[e].y,2s+1$}
                    \State{$Scene.add(sphere(pos=(x,y,z), r=Const))$}
                \EndFor
                \For{$e\in\lbrack 0, E),c\in\lbrack 0, C)$}
                    \State{$res_{c}\leftarrow \lbrack labels[p]=c~for~p~in~res \rbrack$}
                    \State{$Map\lbrack c \rbrack\lbrack e\rbrack\leftarrow \lbrack  p~encoded~by~e~for~p~in~res_{c}\rbrack$}
                \EndFor
            \EndFor
            \State{\textbf{return }$Scene, Map, centroids, labels$}
            \EndFunction
        \end{algorithmic}
    \end{algorithm}

\subsection{Online Searching}
\label{sec:OnlinePhase}
We describe the case of a single query search in the online phase, which is detailed in \Alg{Algo-L2LUT}.
It straightforward to generalize the single-query case to the multi-query case.

Given a query, we first perform the filtering and select $nprobs$ clusters (line 2).
We then calculate the residuals between the query and centroids of the selected clusters (line 4-5).
For dynamic distance threshold, and we calculate the threshold for every query projection with density map, polynomial regressor trained offline and user-defined scaling factor, and transform the threshold to ray's maximal travel time $t_{max}$ (line 6-7).
For each selected cluster, a ray is created from the coordinate of the residual between the query projection and the cluster centroid projection in each subspace (line 8-9). 
As a reminder, the spheres representing the codebook entries in the $s^{th}$ subspace are placed at $z=2s+1$. By placing the ray origin at $z=2s$ and restricting the $t_{max}$ of the rays to be $1.0$, these rays can accurately interact with the spheres in the same subspace without interfering with entries in other subspaces. 
This is illustrated in the left part of \Fig{fig-Intuition}.


Line 10 registers a callback called \texttt{RT\_HitShader}, which would be invoked when the ray hits the sphere, i.e., an entry falls within the interested region of a query projection in a subspace. 
Inside the hit callback, we calculate and record the actual distance between the entry and the query projection using the variable $t_{hit}$ (line 14-16). To process multiple queries, we create and shoot rays of all projections in parallel, maintaining for each query a list that records the hit entry IDs and their corresponding distances in every subspace (line 17-18). These lists collectively form the L2 lookup table.


\begin{algorithm}[b]
    \small
        \caption{Construct L2-LUT with the RT core and conduct distance calculation for interested search points.}
        \label{Algo-L2LUT}
        \textbf{Input:} $queries\lbrack Q \rbrack\lbrack D \rbrack$, $index$, $query\_select\_clusters$, $nprobs$\\
        \textbf{Input:} $density\_map$, $poly\_regressor$, $thres\_scale$(user defined)\\
        \textbf{Output:} $L2\_LUT[Q][nprobs][\frac{D}{M}]\lbrace entry\_id:distance \rbrace$
        \begin{algorithmic}[1]
            \Function {L2\_LUT}{$query$, $index$, $query\_select\_clusters$}
            \For{$q\in\lbrack 0,Q),s\in\lbrack 0, \frac{D}{M})$}
                \State{$x,y,z\leftarrow q\lbrack \_q\rbrack\lbrack 2s:2s+1 \rbrack, 2s$}
                \For{$c$\textbf{ in }$query\_select\_clusters[q]$}
                    \State{$thres\leftarrow poly\_regressor(density\_map(x,y))$}
                    \State{$t\leftarrow 1-\sqrt{1.0^2-(thres\times thres\_scale)^2}$}
                    \State{$x,y\leftarrow (x,y)-index.centroids\lbrack c\rbrack\lbrack 2s:2s+1\rbrack$}
                    \State{$rays.add(x,y,z,t_{max}=t,dir=(0,0,1))$}
                \EndFor
            \EndFor
            \State{$index.scene.set\_hit\_callback($\textbf{RT\_HitShader}$)$}
            \State{\textbf{return }\textbf{RayTracing(}$rays, scene$\textbf{)}}
            \EndFunction
            \Function {RT\_HitShader}{$index$,$query\_select\_clusters$}
            \State{$ray, sphere,t_{hit}\leftarrow$\textbf{GetRay(),GetHitSphere(),GetTime()}}
            \State{$q,s,e\leftarrow ray.query\_id, ray.subspace\_id, sphere.entry\_id$}
            \State{$distance\leftarrow \sqrt{R^2 - (1-t_{hit})^2}$ // $R=Const$}
            \For{$c$\textbf{ in }$query\_select\_clusters$}
                    \State{$L2\_LUT[q][c][s].add(\lbrace e:distance \rbrace)$}
            \EndFor
            \EndFunction
        \end{algorithmic}

    \end{algorithm}

After obtaining the L2-LUT, we perform distance calculations for the search points (formally described in supplemental materials).
For each subspace, we access the inverted index to retrieve the search points whose entry is matched by the query projection, and then accumulate their distances with the results in the L2-LUT. 
The remaining search points would be directly assigned with a large constant without performing any L2-LUT lookup.
Finally, a list containing the search points and their distances are returned for selecting the top-k results.


\begin{figure}[t]
    \centering
      \includegraphics[width=0.485\linewidth]{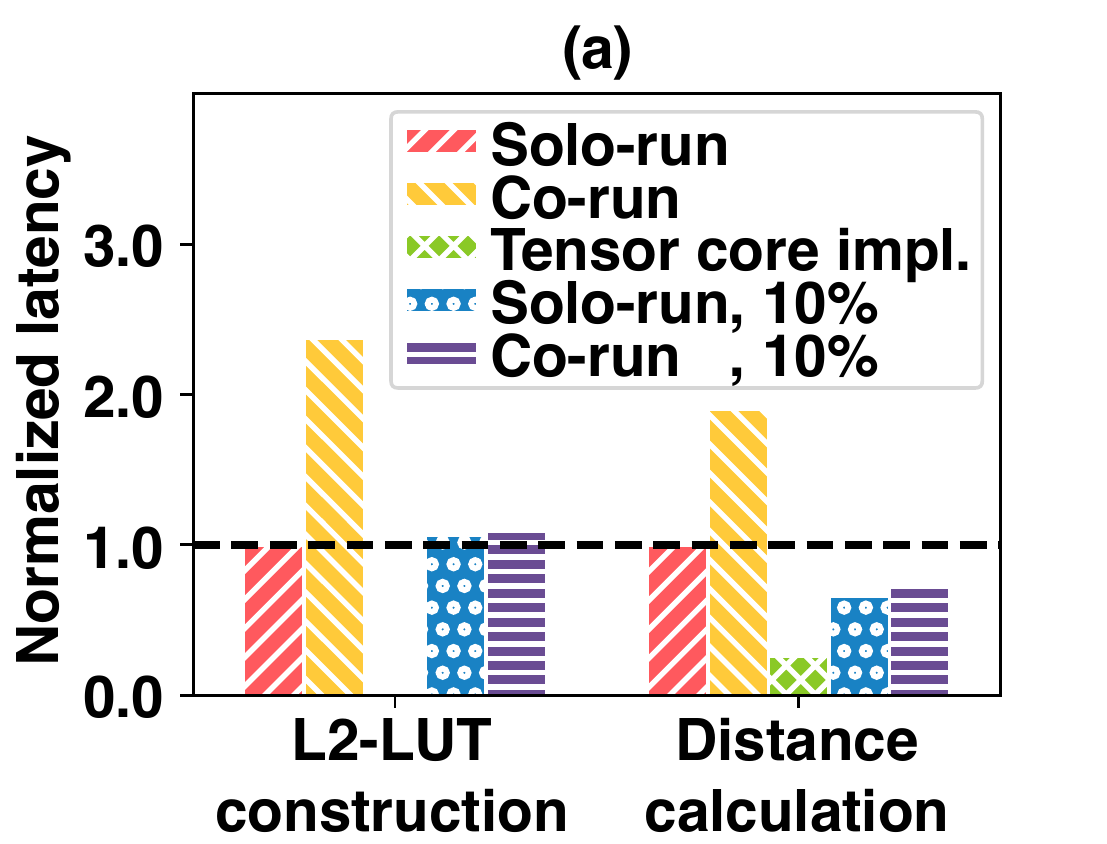}
      \includegraphics[width=0.50\linewidth]{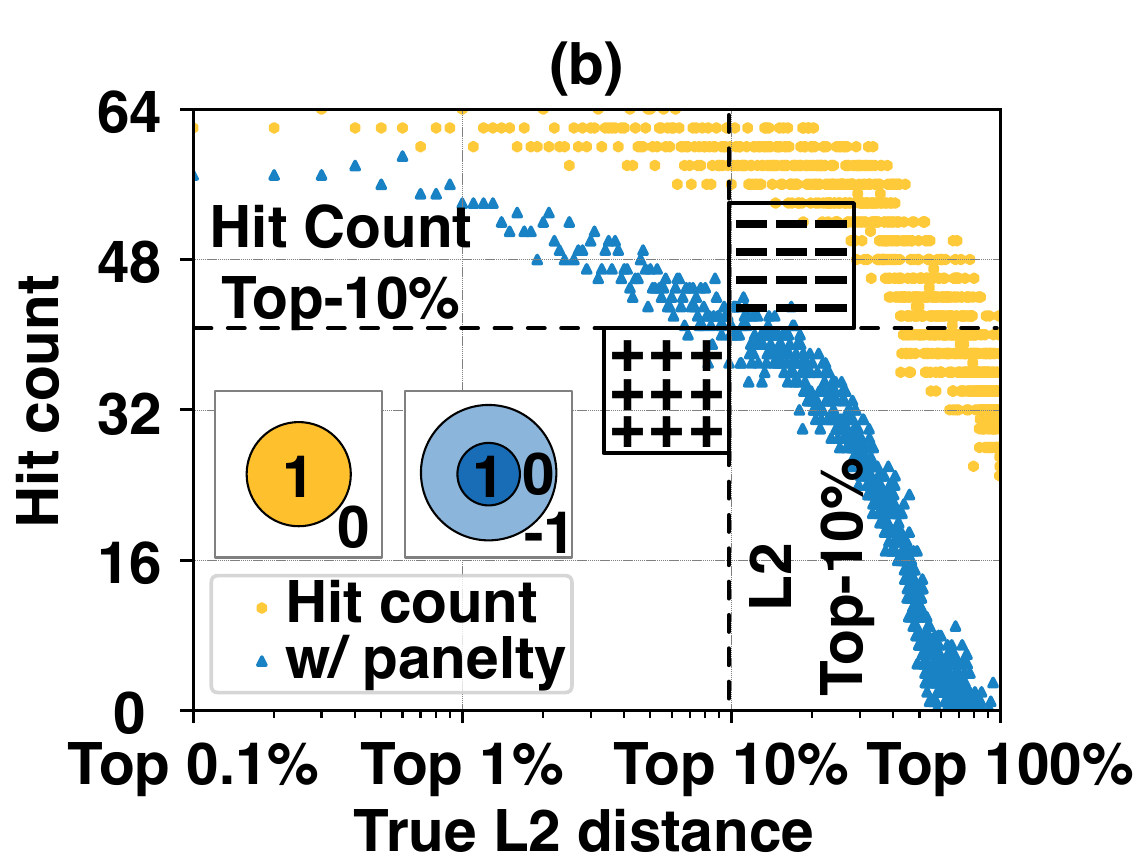}
    \caption{(a) Latency breakdown of three stages. (b) Relationship between hit count and exact distance of query and search points.}
    \label{fig-PipelineAndCounter} 
\end{figure}

\paragraph{Pipelining on Heterogeneous Cores.}
\label{sec:Pipeline}
In our \proj{} framework, the \emph{L2-LUT construction} utilizes the RT core, while the \emph{distance calculation} is performed on the CUDA core. This configuration has the potential for pipelined execution, which can improve the search throughput. This capability is supported by NVIDIA GPUs starting from the Ampere architecture, which allows for co-running of the Tensor core, RT core, and CUDA core~\cite{Ampere}. However, naive co-running without proper optimization can result in severe interference and slowdown, as demonstrated in \Fig{fig-PipelineAndCounter}(a).
The reason is that long latency of distance calculation on the CUDA core leads to severe resource contention.


\proj{} achieves efficient hardware pipelining by mapping the accumulation in the distance calculation stage to Tensor cores. The distances of the selected search points in each subspace are organized into rows to form matrix $A$ with shape $M,K=Q\times \text{sizeof(selected points)}\times nprobs, D/M$ (with padding for simplicity). Then, matrix $B$ is created with shape $K,N=D/M,1$, and all elements are set to $1.0$. The accumulation is then performed by calculating the \texttt{matmul} $A\times B$, utilizing the \texttt{cublas} library~\cite{CUBLAS} on Tensor cores. 
Thus, the reduced latency mitigates the resource contention. 

We utilize CUDA MPS~\cite{MPS} to partition the SM resources in a 9:1 ratio, allocating 90\% of the resources to \emph{L2-LUT construction} (using RT cores) and 10\% to \emph{distance calculation} (using Tensor cores). This partitioning results in similar latencies for the two stages, maximizing the overlap of the two stages. 
We also apply proper data padding and transformation to enable the pipeline, with an overhead of less than 5\% of the latency, compared to the solo-run of \Fig{fig-PipelineAndCounter}(a).


For \emph{filtering}, L2 distance can be calculated with $\lVert \bm{x}-\bm{q}\rVert_{2}^{2}=\bm{x^{2}}-2\bm{x}\bm{q}^{T}+\bm{q^{2}}$, where $\bm{x^2}\leftarrow\sum_{i=0}^{D-1}x_{i}^{2}$ is calculated ahead of time. We just calculate the $\bm{q^2}\leftarrow \sum_{i=0}^{D-1}q_{i}^{2}$ for every query. To calculate $2\bm{x}\bm{q}^{T}$, we use \texttt{cublas} and Tensor core too with $\alpha=-2,\beta=1,A=\bm{x}, B=\bm{q}^{T}, C=\bm{x^{2}}(\bm{q^{2}})^{T}$. Inner product is even simpler by calling a \texttt{matmul} with $A=\bm{x},B=\bm{q}^{T}$.

\subsection{Aggressive Approximation: hit count-based Method}
\label{sec:Aggressive}
The above \emph{L2-LUT construction} still needs to conduct several floating point operations to calculate the radius of a hit sphere and hit distance of a ray.
We propose a more aggressive approximated ANN search that uses the hit/miss result from the RT core, inspired by previous work~\cite{NNHighDim}. 

We first study the relationship between hit count and exact distance,  where all spheres are set with a radius threshold that encompasses the top-100 search points.
As \Fig{fig-PipelineAndCounter}(b) shows, there is strong correlation between these two factors. 
The reason is that higher hit count implies being close to the query projection in more subspaces.
This finding inspires us to implement a hit count-based ANN search method. 

Specifically, we employ a reward/penalty-based model, which involves the extra sphere with half the radius for each original sphere, as in \Fig{fig-PipelineAndCounter}(b).
The hit count is incremented by one only when the ray successfully intersects the inner sphere, and decremented by one as a penalty when the ray misses both spheres. 
As \Fig{fig-PipelineAndCounter}(b) shows, the hit count (blue $\blacktriangle$) calculated using this approach exhibits a stronger correlation compared to the original hit count (yellow $\hexagofill$). While this approximation may result in false positives/negatives (marked as region of `$\bm{-}$' and `$\bm{+}$' respectively in \Fig{fig-PipelineAndCounter}(b)), it offers users a new parameter to trade-off between reasonable search quality degradation and improved throughput.

\begin{figure*}[h]
  \centering
  \includegraphics[width=0.99\linewidth]{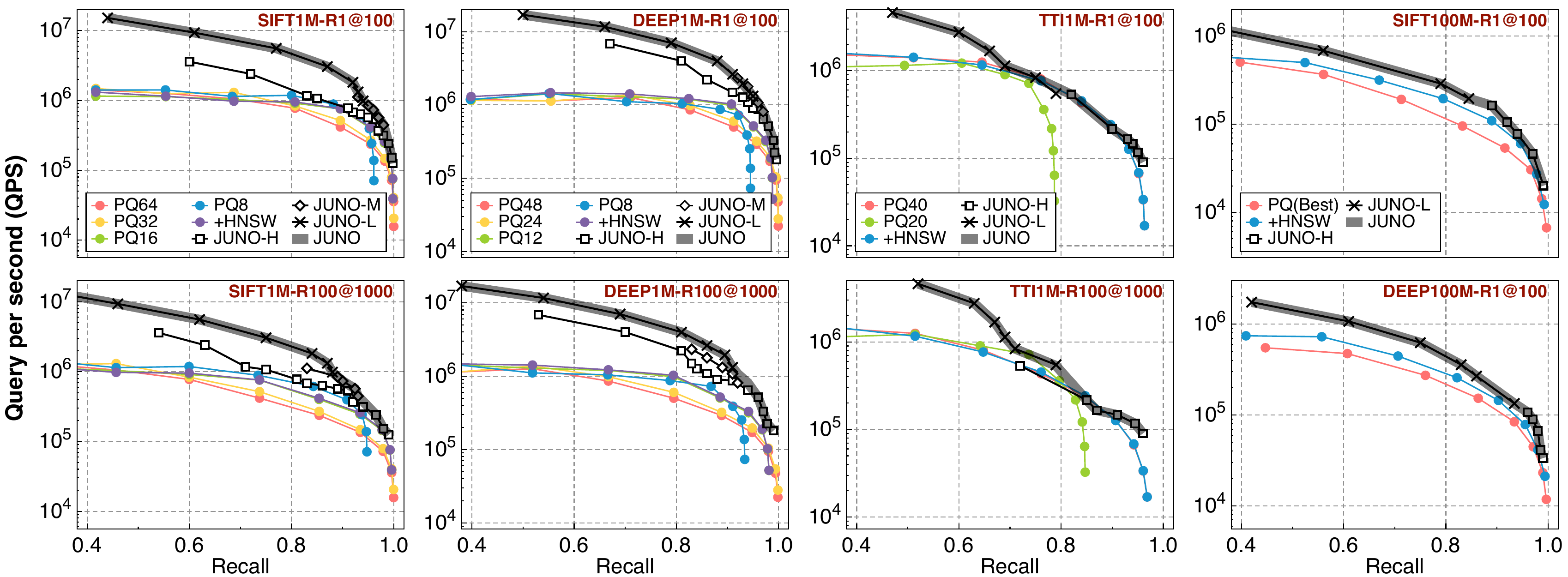}
  \caption{\postrevision{PR-B}{Result of QPS (query per second) and search quality of \proj{} on various datasets including SIFT1M, DEEP1M, TTI1M, SIFT100M and DEEP100M. \revision{R-C}{The bolded grey line labeled \proj{} is the Pareto frontier of our search engine under different configurations (i.e., the configuration of \proj{}-L, \proj{}-M, and \proj{}-H), standing for the optimal performance of \proj{} at a given search quality requirement.}}}
  \label{fig-recallres}
\end{figure*}

\section{Evaulation}
\label{sec:Evaluation}
We demonstrate the effectiveness of the proposed algorithm and hardware mapping of \proj{} throughout experiments.

\subsection{Experimental Setup}
\paragraph{Setup.} 
The ray tracing (RT) part of \proj{} is implemented with NVIDIA OptiX 7.6~\cite{OptiX}.
We evaluate \proj{} on different NVIDIA GPUs (with CUDA/RT cores): RTX 4090 (16384/128), A100 (6912/0), and NVIDIA Tesla A40 (10752/84).
Notice that OptiX offloads the RT computation to CUDA cores if the GPU does not have RT cores. This makes \proj{} also compatible with the GPU without the RT core.
As such, we can conduct the sensitivity analysis to study i) the effectiveness of algorithm enhancement only and ii) the impact of ray tracing hardware performance. 

\paragraph{Dataset.} 
Our work focuses on improving the performance of high-dimensional ANN search on a single GPU.
\revision{R-D}{
We use popular datasets, including SIFT1M/100M~\cite{SIFT1M}, DEEP1M/100M~\cite{DEEP1M} and TTI1M\cite{TTI1M}. The suffix 1M stands for 1 million search points and the suffix 100M stands for 100 million points. Their embedding sizes are 128/128, 96/96 and 200, respectively.}
TTI1M uses the inner product metric (MIPS), and the rest uses the L2 distance metric.
\postrevision{PR-D}{Notice that larger datasets (>1B) cannot be held in the memory of a single GPU, and they need chunking, partitioning, or other storage-related techniques proposed in other orthogonal works~\cite{chen2021spann,DiskANN,iQAN}. For example, GGNN utilizes 8 GPUs, and ANNA utilizes 12 accelerators to process 1B datasets~\cite{GGNN, ANNA}.}
\paragraph{Baseline and Configurations.} 
We mainly compare \proj{} against FAISS, the state-of-the-art GPU-accelerated ANN search library~\cite{FAISS}.
\revision{R-E}{
To comprehensively evaluate the advantages of our work, we use identical \texttt{IVF} cluster numbers as those used in FAISS within \proj{}. Next, we conduct evaluations on various \texttt{PQ} configurations. Finally, we augment our approach by incorporating the widely applied \texttt{HNSW} (Hierarchical Navigable Small World graphs) optimization~\cite{malkov2018efficient} on top of the \texttt{IVF} and \texttt{PQ} methods. It is important to note that \texttt{HNSW} represents an orthogonal index optimization technique, compatible with \texttt{IVF}. 
Since in the search process of \texttt{HNSW}, distance calculation and sorting are still necessary among neighbors of vertices in the navigable small world graphs, and can still benefit from optimized \texttt{PQ} process proposed in \proj{}. 
While our primary focus in \proj{} is on optimizing the \texttt{PQ} component, we defer the integration of \texttt{HNSW} optimization into the current \texttt{IVF} indexing for future research. Nevertheless, we make a thorough comparison of our work against baselines that include \texttt{HNSW} optimizations. In FAISS, this is implemented by calling the \texttt{index\_factory} with the parameter \texttt{IVFx\_HNSWy,PQz}.
}
There are other works, such as CPU-centric~\cite{ScANN}, RAM-centric~\cite{chen2021spann}, and disk-centric~\cite{chen2021spann, DiskANN} ANN search optimization, which are orthogonal to \proj{}.

\paragraph{Metric.} 
\revision{R-F}{In our evaluation, we assess the search quality using two metrics: Recall-1@100 (R1@100) and Recall-100@1000 (R100@1000). 
The R1@100 metric is defined as follows. For a set of Q queries, each with 100 retrieved neighbors, R1@100 represents the count of queries, among the set of Q, where their 100 retrieved neighbors include the true nearest neighbor. It is important to note that R1@100 does not consider the specific order of the 100 retrieved neighbors. Considering a scenario with ten queries. If the retrieved neighbors of eight queries include the true nearest neighbor, then the R1@100 score would be calculated as 8/10.
The R100@1000 metric measures the averaged number of retrieved neighbors, among a total of 1000 retrieved neighbors, that belong to the 100 true nearest neighbors for each query. }
\paragraph{Evaluation Plan.} \revision{R-G}{In this study, we conduct an evaluation of Query Per Second (QPS) and search quality for \proj{} using various configurations. For each configuration, we apply a scaling factor, as referenced in Section \ref{sec:AlgoMap}, to achieve an optimal balance between search quality and system performance.
\begin{itemize}
\item \underline{\proj{}-H}: We employ hit time based exact hit distance calculation for high quality requirement.
\item \underline{\proj{}-M}: We employ finer-grained hit count-based selection with multiple spheres for medium quality requirement.
\item \underline{\proj{}-L}: We employ hit count-based selection only for low quality requirement.
\end{itemize}
It is worth noting that there is some overlap in the search quality among these configurations. And we have empirically divided the search quality requirements into three intervals: $[0.0, 0.95]$, $[0.95, 0.97]$, and $[0.97, 1.0]$. Accordingly, we designate \proj{}-L, \proj{}-M, and \proj{}-H for configurations that correspond to these intervals, respectively. In instances where \proj{}-L or \proj{}-M fails to meet the requirements of the default intervals, we default to using \proj{}-H.
}
We then evaluate the improvement breakdown of two optimization techniques. Finally, we conduct a sensitivity analysis to evaluate the effectiveness of design decisions in \proj{}.


\subsection{Search Quality and Throughput}

\Fig{fig-recallres} shows the overall search quality and throughput on different datasets. 
We aggregate the various configurations of \proj{}-L/M/H into the grey bold line in the plot as \proj{} allows users to make trade-off between search quality and throughput.
The lines labelled with \texttt{PQx} are from the FAISS baseline stand for dividing the entire space into \texttt{x} subspaces. \revision{R-H}{The lines labelled with \texttt{+HNSW} stand for the performance of adding \texttt{HNSW} optimization to the best performed \texttt{PQ} configuration. Notice that the \texttt{HNSW} is configured as the best performed parameter.}

\paragraph{\revision{R-I}{Justifications of Baseline Configurations.}} 
\revision{}{We conduct a detailed analysis on the performance of various baseline methods and provide a rationale for their chosen configurations. Upon investigating the application of \texttt{HNSW} optimization, we observe that it brings limited improvement in smaller datasets (1M) but exhibits more significant enhancements in larger datasets (100M). Moreover, these observations align well with the benchmark results reported in FAISS~\cite{Bench1G}.
Regarding the \texttt{PQ} configuration, we find that incorporating more smaller subspaces results in improved search quality; however, it also leads to lower throughput. We optimize the baseline on different datasets so that the baseline performance approaches closer to the Pareto optimal.
Despite the varying performances of the baseline methods, in our subsequent analysis, we will compare \proj{} against the best-performing one to ensure a fair and rigorous evaluation.}

\paragraph{SIFT1M and DEEP1M}
We achieve $7.8\times$ higher QPS than the baseline in the low search quality scenario, i.e., \proj{}-L with R1@100$\leq 0.95$. 
The improvements are from the exploiting of sparsity and aggressive approximation.
Firstly, by setting a tighter threshold, we can discard more search points, thereby improving the filtering effectiveness. 
Secondly, we perform hit count-based selection without the actual distance calculation.
Additionally, using a tighter threshold leads to fewer hit events due to the reduced size of the spheres. As a result, \proj{} is able to fully exploit the sparsity and spatial locality of the data, which is particularly advantageous when dealing with low search quality requirements.

 
The high requirement for search quality, such as R1@100=0.99, results in increased selected search points. 
Additionally, it is necessary to compute the actual distance to meet this stringent accuracy requirement. These two factors limit the advantages of leveraging sparsity. However, despite these limitations, the \proj{} approach still achieves a throughput improvement of $2.4\times$ compared to the baseline, owing to the tree-based search methodology employed in the RT core.



It is worth noting that \proj{}-L only achieves 0.95 recall for these two datasets as it employs a pure hit count-based approximation approach. 
\proj{}-M is able to improve the search quality to 0.97 by employing the reward/penalty-based approximation approach with extra inner spheres (\Sec{sec:Aggressive}).
The throughput improvement over the baseline is $2.9\times$.
\paragraph{TTI1M} 
This dataset uses the inner product metric (i.e., MIPS).
Since \proj{}-H calculates the exact distance in every subspace, its performance improvement is $2.04\times$,  similar to the two previous datasets with L2 metric.
Notice that the FAISS baseline can also only reach the 0.96 recall. 
While the hit count-based method abandons the $t_{hit}$ information, and intersecting only implies being close in aspect of L2 distance, rather than being similar in aspect of inner product. So, the search quality rapidly drops in dataset with inner product similarity. Thus, the line representing \proj{}-L moves to left. 


\paragraph{\revision{R-J}{SIFT100M and DEEP100M}}
\revision{}{The \proj{}-H and \proj{}-L configurations achieve an averaged improvement of 1.5$\times$ and 2.1$\times$ over the baseline, respectively. The performance gain of \proj{}-H diminishes as the \emph{distance calculation} becomes the bottleneck. 
Notice that these improvements are calculated by comparing \proj{} \textbf{without} \texttt{HNSW} against FAISS \textbf{with} \texttt{HNSW} optimization.
We do not implement \texttt{HNSW} in \proj{} owing to its high code complexity in current FAISS framework.
Meanwhile, implementing it provides no extra insight to guide the optimization of \texttt{PQ}.
Notably, \proj{}-H outperforms the baseline without \texttt{HNSW} optimization by a factor of 3.0$\times$.
}

\paragraph{\revision{R-K}{Results of Different Metrics.}} 
\revision{}{The results for R100@1000 of SIFT1M, DEEP1M, and TTI1M are presented in Figure \ref{fig-recallres}. It is evident that \proj{} exhibits comparable improvements over the baseline, demonstrating the effectiveness of the approximation techniques proposed by \proj{} even under more challenging metrics. Specifically, on average, the top 100 retrieved neighbors out of 1000 contain 65\% of the true top 100 nearest neighbors. This performance is also influenced by the clustering quality of \texttt{IVF} and the \texttt{PQ} quality, which remain identical in \proj{} when compared to the baseline.}

\subsection{Effectiveness of Different Optimizations}

\begin{figure}[b]
    \centering
      \includegraphics[width=0.49\linewidth]{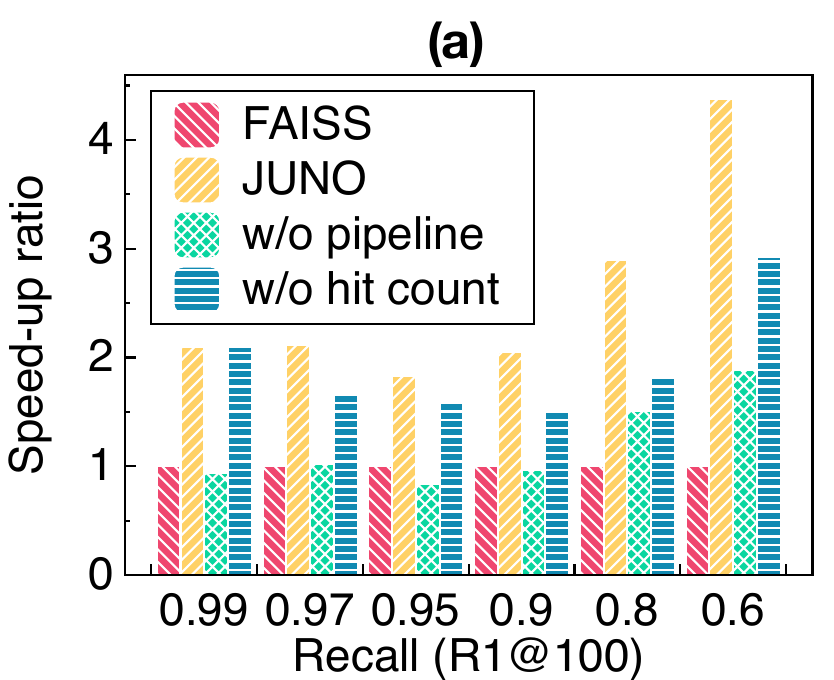}
      \includegraphics[width=0.49\linewidth]{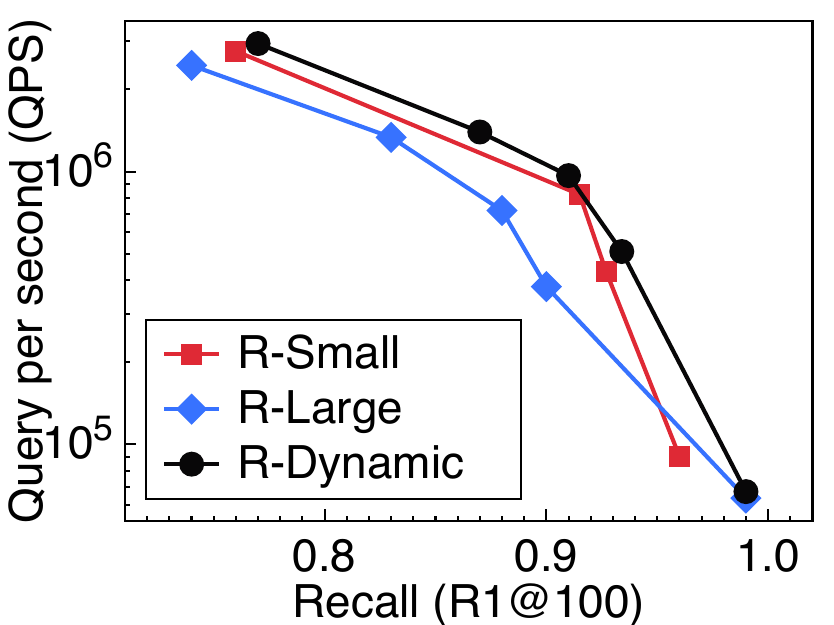}
    \caption{(a) Improvement breakdown of \proj{} against FAISS. (b) Performance of different threshold strategy, evaluated on A40.}
    \label{fig-breakdown} 
\end{figure}

We now evaluate the effectiveness of optimizations used in \proj{}, including the pipelining among CUDA-tensor-RT cores, hit count-based L2-LUT selective construction, and dynamic radius  (i.e., distance threshold).
\Fig{fig-breakdown}(a) shows the both the overall improvement and effects without applying the first two optimizations, while (b) shows the effects of dynamic radius.

\paragraph{Overall Improvement.} 
\revision{R-N}{\proj{} achieves averaged $2.1\times$-$4.4\times$ QPS improvement on five datasets from high to low search quality requirement over the baseline. Although we do not tune \proj{} for a particular dataset, \proj{} still achieves $8.5\times$-$3.2\times$ maximum improvement on these datasets.}

\paragraph{Effectiveness of Pipelining.} 
The third bar in \Fig{fig-breakdown}(a) presents the performance improvement achieved without pipelining. Specifically, in scenarios where high search quality is required, the critical path is the \emph{L2-LUT construction}, which has a longer latency compared to the distance calculation. Consequently, without pipelining, the improvement decreases by 44\%. On the other hand, in situations where lower search quality is acceptable, the latency between the \emph{L2-LUT construction} and \emph{distance calculation} is similar, resulting in a 50\% decrease in the improvement without pipelining.


\paragraph{Effectiveness of Hit Count-based Selection.} 
The final bar in \Fig{fig-breakdown}(a) displays the improvement achieved when hit count-based selection is not utilized. It is evident that for extremely high search quality requirements, hit count-based selection has no influence as it is unable to achieve such high quality. However, as the search quality requirement decreases, the influence of hit count-based selection increases, as lower recall rates necessitate fewer exact distance calculations. In conclusion, a combination of hit count-based selection and exact distance calculation is necessary to deliver high search throughput across different search quality requirements.

\paragraph{Effectiveness of Dynamic Threshold Strategy.} 
We evaluate the search quality and search throughput (QPS) using a small and large static threshold. \revision{R-L}{We conduct this evaluation on SIFT1M dataset with \proj{}-H, where small and large static thresholds are determined with the minimum and maximal threshold values of dynamic threshold.}
As \Fig{fig-breakdown}(b) shows, utilizing a large static threshold results in a decreased search throughput but achieves higher search quality. This can be attributed to the fact that a larger threshold leads to a larger sphere, causing a ray to intersect with more spheres and trigger more hit shader functions. 
Conversely, using a small static threshold improves the search throughput, but at the cost of degraded search quality. 
Even for scenarios with low search quality requirements, selecting more clusters to address the recall issue becomes necessary, thereby negating the potential performance gain from reducing hit shader invocations. For cases with high search quality requirements, this small static threshold proves to be inadequate, as it results in missing too many true top-k neighbors, thereby jeopardizing recall. In contrast, our dynamic threshold strategy outperforms both the small and large static threshold-based approaches in terms of both search quality and throughput.




\subsection{Sensitivity to RT Core Performance}

\begin{figure}[t]
    \centering
        \includegraphics[width=0.49\linewidth]{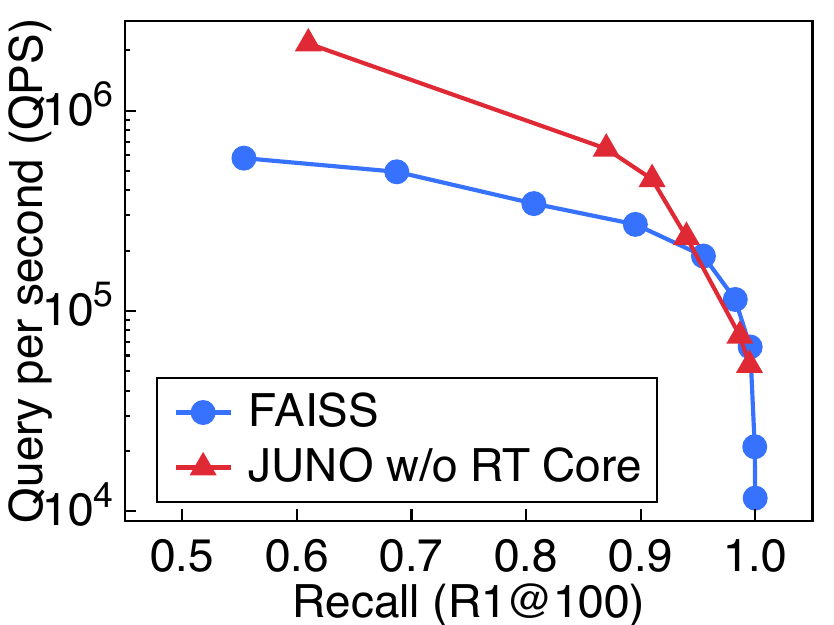}
      \includegraphics[width=0.49\linewidth]{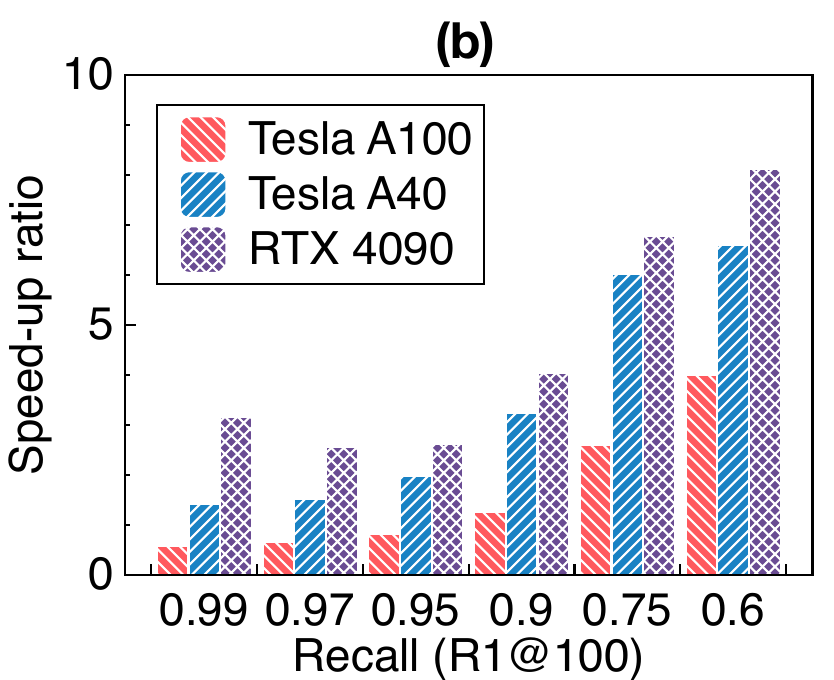}
    \caption{(a) QPS and recall of \proj{} and FAISS on A100. (b) Average advantage against FAISS on different GPUs.}
    \label{fig-sensitivity} 
\end{figure}

Finally, we evaluate the performance of \proj{} with and without the acceleration of the RT core using different GPUs.
\Fig{fig-sensitivity}(a) shows the detailed performance of \proj{} and the baseline on Tesla A100, a GPU without the RT core. \revision{R-M}{We conduct this evaluation on SIFT1M dataset. The baseline is configured with \texttt{PQ16+HNSW} (the best performed configuration among others).} We find that \proj{} still gains significant improvement with low search quality requirement on Tesla A100, which means the advantage is purely contributed by the threshold-based selective algorithm. This result also verifies that it is reasonable to leverage the sparsity and spatial similarity in the typical \texttt{IVFPQ} process. For high search quality requirements, \proj{} gradually performs worse than the baseline since the overhead to simulate ray tracing with the CUDA core suppresses the tiny positive effect brought by the sparsity.


Results in \Fig{fig-PipelineAndCounter}(a) imply that the performance of \proj{} is bound by the performance of RT cores.
So, we expect a performance improvement with a faster RT core. According to the white paper of NVIDIA Ada architecture~\cite{Ada}, the Gen.3 RT core of Ada GPUs has 2$\times$ throughput compared to a Gen.2 RT core of Ampere GPUs. As shown in \Fig{fig-sensitivity}(b), \revision{R-N}{on average of three 1M datasets,} RTX4090 has 1.5$\times$ higher improvement over the baseline than Tesla A40. 
Notice that the throughput of CUDA cores and Tensor cores of RTX4090 is 1.4$\times$ of A40 GPU per SM~\cite{Ampere,Ada}.

\subsection{Robustness Discussions}
\revision{R-R}{Since \proj{} adopts an approximate search methodology, we delve deeper into its robustness aspect, focusing on concerns that researchers may have pertaining to accuracy guarantees and dataset characteristics.
\paragraph{Accuracy Guarantee.}
Similar to our baseline ANN search framework, \proj{} does not offer a theoretical accuracy guarantee and relies on the quality of offline clustering. Nonetheless, it has the capability to support lossless searching through a series of minor adjustments:
i) exhaustively searching all \texttt{IVF} clusters, equivalent to a \texttt{Flat} index, ensures that all search points are thoroughly explored, leaving no potential neighbors missed.
ii) projecting all original search points (rather than PQ codebook entries) into a two-dimensional space,
iii) and employing ray tracing to calculate precise distances between search points (rather than PQ codebook entries) and queries.
As a result, we can guarantee the search accuracy and use 
\proj{} in scenarios with stringent search quality requirements.}

\begin{figure}[h]
  \centering
  \includegraphics[width=0.99\linewidth]{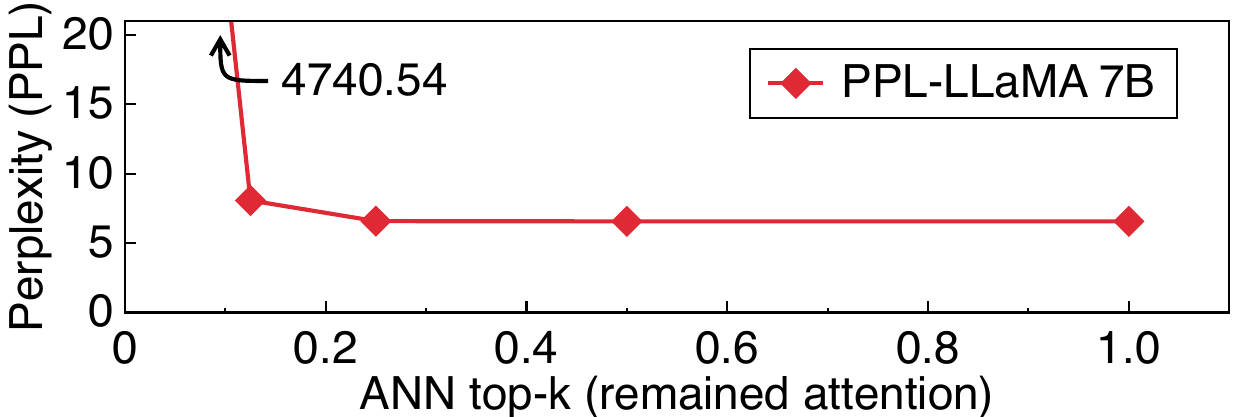}
  \caption{\revision{}{Word perplexity of Llama-7B with different amount of attention remained.}}
  \label{fig-TopkLLM} 
\end{figure}

\paragraph{\revision{}{Dataset Characteristics.}}
\revision{}{\proj{} leverages sparsity and locality inherent in high-dimensional ANN search to improve efficiency. Critically, \proj{} does NO dataset-specific tuning and optimization. As we have shown in the above results, \proj{} consistently delivers significant performance improvements across a number of popular datasets, despite their different levels of sparsity and redundancy characteristics.}
\postrevision{PR-C}{
Furthermore, in the era of training expansive foundational models using vast datasets, the data exhibit an escalating trend of becoming increasingly sparse and redundant~\cite{SparseGPT, LotteryTicket, SurveyOnPruning, SurveyOnSparsity, TileWiseSparsity}. 
Those foundational models adopt the decoder-based Transformer architecture that adopts the multi-head attention mechanism.  
The memory and computation complexity of attention mechanism scales quadraticly with the sequence length, making it the main bottleneck when dealing with long input sequences.
On the other hand, the attention mechanism essentially calculates the inner product between query vectors and key vectors, which correspond to the query points and search points in vector search.
Prior works have shown that keeping the most significant attention values can still preserve the model accuracy~\cite{Reformer,Unlimiformer}, making LLM an ideal candidate to be accelerated by ANN search. This phenomenon underscores the promising potential of JUNO in future.\\
To verify, we conduct an experiment on Llama-7B~\cite{LLAMA7B}. Recall that 30\%-50\% nearest entries should be preserved to maintain search qualities in ANN search according to \Fig{fig-Motivation} and \Fig{fig-DSSparsityCDF}, while as shown in \Fig{fig-TopkLLM}, a commendable quality can be sustained with less than 20\% nearest tokens attended.}

\section{Related Work}
\label{sec:RelatedWorks}
Since there are two aspects, i.e., indexing and encoding, in an ANN algorithm, we compare \proj{} with existing works in these two algorithmic aspects and hardware acceleration. 

\paragraph{Indexing.}
Indexing techniques aim to compress the search space by eliminating unnecessary search points. 
Except for the trival flat index that store the complete database, one example of such technique is the inverted file index (\texttt{IVF})~\cite{IVF}, which organizes search points into clusters and selects several closest clusters when searching (search points in other unselected clusters are ignored). 
Another type of technique involves graph-based methods that construct a nearest neighbor graph to quickly prune the search space~\cite{dong2011efficient,hajebi2011fast,wang2012scalable}. 
To accelerate these methods, heuristic-based approaches~\cite{baranchuk2019learning,fu2017fast,iwasaki2018optimization} have been proposed. 
\revision{R-O}{
Currently, among these works, hierarchical navigable small world (\texttt{HNSW})~\cite{malkov2018efficient} and navigating spread-out graph (\texttt{NSG})~\cite{NSG} are the most representative. 
The \texttt{HNSW} construct the neighbor graph hierarchically, with the search going deeper, the graph have higher degree and shorter edges. So that the search can quickly choose a good search direction and get good enough results in log-scale. 
The \texttt{NSG} further reduce the size and degree of graphs, thus reduce the length of search path via setting navigation points from which the search begins. 
}
Additionally, tree-based techniques including kd-tree and octree~\cite{muja2014scalable,ParallelNN}, and locality-sensitive hashing (\texttt{LSH})~\cite{datar2004locality,dasgupta2011fast} are also commonly used for indexing.
\revision{R-P}{
Noted that \proj{} is not limited to specific indexing methods, and is compatible with \texttt{Flat}, \texttt{IVF}, \texttt{HNSW}, etc. 
}

\paragraph{Encoding.}
Encoding techniques aim to reduce the memory consumption of search points. The most commonly used technique is product quantization (\texttt{PQ})~\cite{jegou2010product}, which splits the space into several subspaces and encodes the search point projections. Several techniques have been proposed to optimize the codebook quality thanks to its strong relation between search quality, such as \texttt{DPQ}~\cite{klein2019end} and \texttt{OPQ}~\cite{ge2013optimized}. Additionally, scalar quantization (\texttt{SQ})~\cite{SQ} maps vector components separately and linearly, similar to traditional quantization in DNNs~\cite{ANT, OliVe}. Additive quantization (\texttt{AQ})~\cite{AQ} encodes search points as a sum of codebook entries.
\revision{R-P}{
\proj{} currently supports product quantization (\texttt{PQ}) only.
}


\paragraph{Hardware Acceleration.}
There are several specialized architecture designs including hardware support for hierarchical product quantization~\cite{abdelhadi2019accelerated} and fused high-performance k-selection~\cite{zhang2018efficient}. ANNA proposed an end-to-end hardware solution for PQ-based ANN search~\cite{ANNA}. 
Several architectural designs based on tree-like data structures~\cite{Tigris, ParallelNN} are proposed for low-dimensional ANN search.
In addition to computation, large-scale ANN search presents severe challenges to the memory and storage subsystems. DiskANN presented a graph-based indexing that can search with limited RAM and cheap solid-state drives~\cite{jayaram2019diskann}. SPANN presented a memory-disk hybrid indexing following the inverted file index~\cite{chen2021spann}.
Those solutions employ partitioning techniques to divide large datasets into smaller chunks for index creation and processing, in order to overcome resource limitations. Particularly in the case of GPU-accelerated ANNS, the GPU's memory capacity is often inadequate to load the entire dataset. As such, we believe that evaluating a 100M dataset would be adequate to showcase the superior performance of JUNO.

Researchers have also proposed various techniques to map the ANN search to existing hardware. ScANN optimizes the ANN search with the AVX ISA on CPUs~\cite{ScANN}. RTNN maps the low-dimensional ANN search to the RT core~\cite{RTNN}. 
Our system seeks to harness the power of the RT core to accelerate more general ANN search in high-dimensional spaces.


\section{Conclusion}
\label{sec:Conclusion}

In this work, we have presented \proj{}, an end-to-end approximate nearest neighbor (ANN) search system that incorporates a sparsity-aware codebook entry selection algorithm and a highly efficient RT core mapping.
The key in our algorithm is to exploit the opportunities of sparsity and spatial locality that we have identified through detailed profiling.
Specifically, we employ a distance threshold filtering that can be efficiently mapped to RT cores.
Additionally, we optimize the system with time-based hit distance calculation, hit count-based aggressive approximation, and Tensor-RT core pipelining. 
Evaluation of \proj{} on multiples datasets demonstrates a 2.1$\times$-8.5$\times$ improvement over existing product quantization based ANN search in search throughput across various scenarios.



\section*{Acknowledgement}
This work was supported by the National Key R\&D Program of China under Grant 2022YFB4501401, the National Natural Science Foundation of China (NSFC) grant (62222210, and 62072297, and U21B2017).
We would like to thank the anonymous reviewers for their constructive feedback and comments to improve this work. We also thank our shepherd, Professor Sasa Misailovic, for his support and guidance in preparing this paper for publication. We also thank Yue Guan, Yangjie Zhou, Weihao Cui, Sean Yao, Tinghan Qian, Yingzhe Lyu, and many other colleagues for beneficial discussion and useful comments. Special thanks to Vega Jiang for continuous help and support. Any opinions, findings, and conclusions in this paper are those of the authors only and do not necessarily reflect the views of our sponsors.

\newpage
\balance
\bibliographystyle{plain}
\bibliography{references}

\begin{thebibliography}{10}

\bibitem{abdelhadi2019accelerated}
Ameer~MS Abdelhadi, Christos-Savvas Bouganis, and George~A Constantinides.
\newblock Accelerated approximate nearest neighbors search through hierarchical
  product quantization.
\newblock In {\em 2019 International Conference on Field-Programmable
  Technology (ICFPT)}, pages 90--98. IEEE, 2019.

\bibitem{KNNDiffusion}
Oron Ashual, Shelly Sheynin, Adam Polyak, Uriel Singer, Oran Gafni, Eliya
  Nachmani, and Yaniv Taigman.
\newblock Knn-diffusion: Image generation via large-scale retrieval.
\newblock {\em CoRR}, abs/2204.02849, 2022.

\bibitem{AQ}
Artem Babenko and Victor~S. Lempitsky.
\newblock Additive quantization for extreme vector compression.
\newblock In {\em 2014 {IEEE} Conference on Computer Vision and Pattern
  Recognition, {CVPR} 2014, Columbus, OH, USA, June 23-28, 2014}, pages
  931--938. {IEEE} Computer Society, 2014.

\bibitem{DEEP1M}
Artem Babenko and Victor~S. Lempitsky.
\newblock Efficient indexing of billion-scale datasets of deep descriptors.
\newblock In {\em 2016 {IEEE} Conference on Computer Vision and Pattern
  Recognition, {CVPR} 2016, Las Vegas, NV, USA, June 27-30, 2016}, pages
  2055--2063. {IEEE} Computer Society, 2016.

\bibitem{L2IP1}
Yoram Bachrach, Yehuda Finkelstein, Ran Gilad{-}Bachrach, Liran Katzir, Noam
  Koenigstein, Nir Nice, and Ulrich Paquet.
\newblock Speeding up the xbox recommender system using a euclidean
  transformation for inner-product spaces.
\newblock In Alfred Kobsa, Michelle~X. Zhou, Martin Ester, and Yehuda Koren,
  editors, {\em Eighth {ACM} Conference on Recommender Systems, RecSys '14,
  Foster City, Silicon Valley, CA, {USA} - October 06 - 10, 2014}, pages
  257--264. {ACM}, 2014.

\bibitem{baranchuk2019learning}
Dmitry Baranchuk, Dmitry Persiyanov, Anton Sinitsin, and Artem Babenko.
\newblock Learning to route in similarity graphs.
\newblock In {\em International Conference on Machine Learning}, pages
  475--484. PMLR, 2019.

\bibitem{CVPRANN}
Jeffrey~S. Beis and David~G. Lowe.
\newblock Shape indexing using approximate nearest-neighbour search in
  high-dimensional spaces.
\newblock In {\em 1997 Conference on Computer Vision and Pattern Recognition
  {(CVPR} '97), June 17-19, 1997, San Juan, Puerto Rico}, pages 1000--1006.
  {IEEE} Computer Society, 1997.

\bibitem{Unlimiformer}
Amanda Bertsch, Uri Alon, Graham Neubig, and Matthew~R. Gormley.
\newblock Unlimiformer: Long-range transformers with unlimited length input.
\newblock {\em CoRR}, abs/2305.01625, 2023.

\bibitem{ParallelNN}
Faquan Chen, Rendong Ying, Jianwei Xue, Fei Wen, and Peilin Liu.
\newblock Parallelnn: {A} parallel octree-based nearest neighbor search
  accelerator for 3d point clouds.
\newblock In {\em {IEEE} International Symposium on High-Performance Computer
  Architecture, {HPCA} 2023, Montreal, QC, Canada, February 25 - March 1,
  2023}, pages 403--414. {IEEE}, 2023.

\bibitem{chen2021spann}
Qi~Chen, Bing Zhao, Haidong Wang, Mingqin Li, Chuanjie Liu, Zengzhong Li, Mao
  Yang, and Jingdong Wang.
\newblock Spann: Highly-efficient billion-scale approximate nearest
  neighborhood search.
\newblock {\em Advances in Neural Information Processing Systems},
  34:5199--5212, 2021.

\bibitem{RecommendationANN}
Rihan Chen, Bin Liu, Han Zhu, Yaoxuan Wang, Qi~Li, Buting Ma, Qingbo Hua, Jun
  Jiang, Yunlong Xu, Hongbo Deng, and Bo~Zheng.
\newblock Approximate nearest neighbor search under neural similarity metric
  for large-scale recommendation.
\newblock In Mohammad~Al Hasan and Li~Xiong, editors, {\em Proceedings of the
  31st {ACM} International Conference on Information {\&} Knowledge Management,
  Atlanta, GA, USA, October 17-21, 2022}, pages 3013--3022. {ACM}, 2022.

\bibitem{SurveyOnPruning}
Hongrong Cheng, Miao Zhang, and Javen~Qinfeng Shi.
\newblock A survey on deep neural network pruning-taxonomy, comparison,
  analysis, and recommendations.
\newblock {\em CoRR}, abs/2308.06767, 2023.

\bibitem{dasgupta2011fast}
Anirban Dasgupta, Ravi Kumar, and Tam{\'a}s Sarl{\'o}s.
\newblock Fast locality-sensitive hashing.
\newblock In {\em Proceedings of the 17th ACM SIGKDD international conference
  on Knowledge discovery and data mining}, pages 1073--1081, 2011.

\bibitem{datar2004locality}
Mayur Datar, Nicole Immorlica, Piotr Indyk, and Vahab~S Mirrokni.
\newblock Locality-sensitive hashing scheme based on p-stable distributions.
\newblock In {\em Proceedings of the twentieth annual symposium on
  Computational geometry}, pages 253--262, 2004.

\bibitem{ROI}
Min Dong, Zhe Wang, Chenghui Dong, Xiaomin Mu, and Yide Ma.
\newblock Classification of region of interest in mammograms using dual
  contourlet transform and improved {KNN}.
\newblock {\em J. Sensors}, 2017:3213680:1--3213680:15, 2017.

\bibitem{dong2011efficient}
Wei Dong, Charikar Moses, and Kai Li.
\newblock Efficient k-nearest neighbor graph construction for generic
  similarity measures.
\newblock In {\em Proceedings of the 20th international conference on World
  wide web}, pages 577--586, 2011.

\bibitem{octree}
J.~Elseberg, S.~Magnenat, R.~Siegwart, and A.~N{\"u}chter.
\newblock Comparison of nearest-neighbor-search strategies and implementations
  for efficient shape registration.
\newblock {\em Journal of Software Engineering for Robotics (JOSER)},
  3(1):2--12, 2012.

\bibitem{LotteryTicket}
Jonathan Frankle and Michael Carbin.
\newblock The lottery ticket hypothesis: Finding sparse, trainable neural
  networks.
\newblock In {\em 7th International Conference on Learning Representations,
  {ICLR} 2019, New Orleans, LA, USA, May 6-9, 2019}. OpenReview.net, 2019.

\bibitem{SparseGPT}
Elias Frantar and Dan Alistarh.
\newblock Sparsegpt: Massive language models can be accurately pruned in
  one-shot.
\newblock In Andreas Krause, Emma Brunskill, Kyunghyun Cho, Barbara Engelhardt,
  Sivan Sabato, and Jonathan Scarlett, editors, {\em International Conference
  on Machine Learning, {ICML} 2023, 23-29 July 2023, Honolulu, Hawaii, {USA}},
  volume 202 of {\em Proceedings of Machine Learning Research}, pages
  10323--10337. {PMLR}, 2023.

\bibitem{fu2017fast}
Cong Fu, Chao Xiang, Changxu Wang, and Deng Cai.
\newblock Fast approximate nearest neighbor search with the navigating
  spreading-out graph.
\newblock {\em arXiv preprint arXiv:1707.00143}, 2017.

\bibitem{NSG}
Cong Fu, Chao Xiang, Changxu Wang, and Deng Cai.
\newblock Fast approximate nearest neighbor search with the navigating
  spreading-out graph.
\newblock {\em Proc. VLDB Endow.}, 12(5):461–474, jan 2019.

\bibitem{LinearANN}
Jianyang Gao and Cheng Long.
\newblock High-dimensional approximate nearest neighbor search: with reliable
  and efficient distance comparison operations.
\newblock {\em CoRR}, abs/2303.09855, 2023.

\bibitem{ge2013optimized}
Tiezheng Ge, Kaiming He, Qifa Ke, and Jian Sun.
\newblock Optimized product quantization.
\newblock {\em IEEE transactions on pattern analysis and machine intelligence},
  36(4):744--755, 2013.

\bibitem{GGNN}
Fabian Groh, Lukas Ruppert, Patrick Wieschollek, and Hendrik P.~A. Lensch.
\newblock {GGNN:} graph-based {GPU} nearest neighbor search.
\newblock {\em {IEEE} Trans. Big Data}, 9(1):267--279, 2023.

\bibitem{TileWiseSparsity}
Cong Guo, Bo~Yang Hsueh, Jingwen Leng, Yuxian Qiu, Yue Guan, Zehuan Wang,
  Xiaoying Jia, Xipeng Li, Minyi Guo, and Yuhao Zhu.
\newblock Accelerating sparse {DNN} models without hardware-support via
  tile-wise sparsity.
\newblock In Christine Cuicchi, Irene Qualters, and William~T. Kramer, editors,
  {\em Proceedings of the International Conference for High Performance
  Computing, Networking, Storage and Analysis, {SC} 2020, Virtual Event /
  Atlanta, Georgia, USA, November 9-19, 2020}, page~16. {IEEE/ACM}, 2020.

\bibitem{OliVe}
Cong Guo, Jiaming Tang, Weiming Hu, Jingwen Leng, Chen Zhang, Fan Yang, Yunxin
  Liu, Minyi Guo, and Yuhao Zhu.
\newblock Olive: Accelerating large language models via hardware-friendly
  outlier-victim pair quantization.
\newblock In Yan Solihin and Mark~A. Heinrich, editors, {\em Proceedings of the
  50th Annual International Symposium on Computer Architecture, {ISCA} 2023,
  Orlando, FL, USA, June 17-21, 2023}, pages 3:1--3:15. {ACM}, 2023.

\bibitem{ANT}
Cong Guo, Chen Zhang, Jingwen Leng, Zihan Liu, Fan Yang, Yunxin Liu, Minyi Guo,
  and Yuhao Zhu.
\newblock {ANT:} exploiting adaptive numerical data type for low-bit deep
  neural network quantization.
\newblock In {\em 55th {IEEE/ACM} International Symposium on Microarchitecture,
  {MICRO} 2022, Chicago, IL, USA, October 1-5, 2022}, pages 1414--1433. {IEEE},
  2022.

\bibitem{ScANN}
Ruiqi Guo, Philip Sun, Erik Lindgren, Quan Geng, David Simcha, Felix Chern, and
  Sanjiv Kumar.
\newblock Accelerating large-scale inference with anisotropic vector
  quantization.
\newblock In {\em Proceedings of the 37th International Conference on Machine
  Learning, {ICML} 2020, 13-18 July 2020, Virtual Event}, volume 119 of {\em
  Proceedings of Machine Learning Research}, pages 3887--3896. {PMLR}, 2020.

\bibitem{Gems}
Eric Haines and Tomas Akenine-M\"oller, editors.
\newblock {\em Ray Tracing Gems}.
\newblock Apress, 2019.
\newblock \url{http://raytracinggems.com}.

\bibitem{hajebi2011fast}
Kiana Hajebi, Yasin Abbasi-Yadkori, Hossein Shahbazi, and Hong Zhang.
\newblock Fast approximate nearest-neighbor search with k-nearest neighbor
  graph.
\newblock In {\em Twenty-Second International Joint Conference on Artificial
  Intelligence}, 2011.

\bibitem{SurveyOnSparsity}
Torsten Hoefler, Dan Alistarh, Tal Ben{-}Nun, Nikoli Dryden, and Alexandra
  Peste.
\newblock Sparsity in deep learning: Pruning and growth for efficient inference
  and training in neural networks.
\newblock {\em J. Mach. Learn. Res.}, 22:241:1--241:124, 2021.

\bibitem{iwasaki2018optimization}
Masajiro Iwasaki and Daisuke Miyazaki.
\newblock Optimization of indexing based on k-nearest neighbor graph for
  proximity search in high-dimensional data.
\newblock {\em arXiv preprint arXiv:1810.07355}, 2018.

\bibitem{jayaram2019diskann}
Suhas Jayaram~Subramanya, Fnu Devvrit, Harsha~Vardhan Simhadri, Ravishankar
  Krishnawamy, and Rohan Kadekodi.
\newblock Diskann: Fast accurate billion-point nearest neighbor search on a
  single node.
\newblock {\em Advances in Neural Information Processing Systems}, 32, 2019.

\bibitem{jegou2010product}
Herve Jegou, Matthijs Douze, and Cordelia Schmid.
\newblock Product quantization for nearest neighbor search.
\newblock {\em IEEE transactions on pattern analysis and machine intelligence},
  33(1):117--128, 2010.

\bibitem{SIFT1M}
Herv{\'{e}} J{\'{e}}gou, Romain Tavenard, Matthijs Douze, and Laurent Amsaleg.
\newblock Searching in one billion vectors: Re-rank with source coding.
\newblock In {\em Proceedings of the {IEEE} International Conference on
  Acoustics, Speech, and Signal Processing, {ICASSP} 2011, May 22-27, 2011,
  Prague Congress Center, Prague, Czech Republic}, pages 861--864. {IEEE},
  2011.

\bibitem{FAISS}
Jeff Johnson, Matthijs Douze, and Herv{\'e} J{\'e}gou.
\newblock Billion-scale similarity search with {GPUs}.
\newblock {\em IEEE Transactions on Big Data}, 7(3):535--547, 2019.

\bibitem{Reformer}
Nikita Kitaev, Lukasz Kaiser, and Anselm Levskaya.
\newblock Reformer: The efficient transformer.
\newblock In {\em 8th International Conference on Learning Representations,
  {ICLR} 2020, Addis Ababa, Ethiopia, April 26-30, 2020}. OpenReview.net, 2020.

\bibitem{klein2019end}
Benjamin Klein and Lior Wolf.
\newblock End-to-end supervised product quantization for image search and
  retrieval.
\newblock In {\em Proceedings of the IEEE/CVF Conference on Computer Vision and
  Pattern Recognition}, pages 5041--5050, 2019.

\bibitem{NNHighDim}
Jon~M. Kleinberg.
\newblock Two algorithms for nearest-neighbor search in high dimensions.
\newblock In Frank~Thomson Leighton and Peter~W. Shor, editors, {\em
  Proceedings of the Twenty-Ninth Annual {ACM} Symposium on the Theory of
  Computing, El Paso, Texas, USA, May 4-6, 1997}, pages 599--608. {ACM}, 1997.

\bibitem{ANNA}
Yejin Lee, Hyunji Choi, Sunhong Min, Hyunseung Lee, Sangwon Beak, Dawoon Jeong,
  Jae~W. Lee, and Tae~Jun Ham.
\newblock {ANNA:} specialized architecture for approximate nearest neighbor
  search.
\newblock In {\em {IEEE} International Symposium on High-Performance Computer
  Architecture, {HPCA} 2022, Seoul, South Korea, April 2-6, 2022}, pages
  169--183. {IEEE}, 2022.

\bibitem{EarlyQuit}
Conglong Li, Minjia Zhang, David~G. Andersen, and Yuxiong He.
\newblock Improving approximate nearest neighbor search through learned
  adaptive early termination.
\newblock In David Maier, Rachel Pottinger, AnHai Doan, Wang{-}Chiew Tan,
  Abdussalam Alawini, and Hung~Q. Ngo, editors, {\em Proceedings of the 2020
  International Conference on Management of Data, {SIGMOD} Conference 2020,
  online conference [Portland, OR, USA], June 14-19, 2020}, pages 2539--2554.
  {ACM}, 2020.

\bibitem{IVF}
Yuchen Liu, Zhibin Pan, Liangzhuang Wang, and Yang Wang.
\newblock A new fast inverted file-based algorithm for approximate nearest
  neighbor search without accuracy reduction.
\newblock {\em Inf. Sci.}, 608:613--629, 2022.

\bibitem{malkov2018efficient}
Yu~A Malkov and Dmitry~A Yashunin.
\newblock Efficient and robust approximate nearest neighbor search using
  hierarchical navigable small world graphs.
\newblock {\em IEEE transactions on pattern analysis and machine intelligence},
  42(4):824--836, 2018.

\bibitem{Gems2}
Adam Marrs, Peter Shirley, , and Ingo Wald, editors.
\newblock {\em Ray Tracing Gems II}.
\newblock Apress, 2021.
\newblock \url{http://raytracinggems.com/rtg2}.

\bibitem{Bench1G}
Meta.
\newblock Indexing 1g vectors.
\newblock
  \url{https://github.com/facebookresearch/faiss/wiki/Indexing-1G-vectors},
  2023.

\bibitem{muja2014scalable}
Marius Muja and David~G Lowe.
\newblock Scalable nearest neighbor algorithms for high dimensional data.
\newblock {\em IEEE transactions on pattern analysis and machine intelligence},
  36(11):2227--2240, 2014.

\bibitem{DLRM}
Maxim Naumov, Dheevatsa Mudigere, Hao{-}Jun~Michael Shi, Jianyu Huang,
  Narayanan Sundaraman, Jongsoo Park, Xiaodong Wang, Udit Gupta, Carole{-}Jean
  Wu, Alisson~G. Azzolini, Dmytro Dzhulgakov, Andrey Mallevich, Ilia
  Cherniavskii, Yinghai Lu, Raghuraman Krishnamoorthi, Ansha Yu, Volodymyr
  Kondratenko, Stephanie Pereira, Xianjie Chen, Wenlin Chen, Vijay Rao, Bill
  Jia, Liang Xiong, and Misha Smelyanskiy.
\newblock Deep learning recommendation model for personalization and
  recommendation systems.
\newblock {\em CoRR}, abs/1906.00091, 2019.

\bibitem{OptiX}
NVIDIA.
\newblock Nvidia optix™ ray tracing engine.
\newblock \url{https://developer.nvidia.com/rtx/ray-tracing/optix}.

\bibitem{Turing}
NVIDIA.
\newblock Nvidia turing gpu architecture.
\newblock
  \url{https://images.nvidia.com/aem-dam/en-zz/Solutions/design-visualization/technologies/turing-architecture/NVIDIA-Turing-Architecture-Whitepaper.pdf},
  2018.

\bibitem{Ampere}
NVIDIA.
\newblock Nvidia ampere ga102 gpu architecture.
\newblock
  \url{https://www.nvidia.com/content/PDF/nvidia-ampere-ga-102-gpu-architecture-whitepaper-v2.pdf},
  2021.

\bibitem{MPS}
NVIDIA.
\newblock Multi-process service.
\newblock
  \url{https://docs.nvidia.com/deploy/pdf/CUDA_Multi_Process_Service_Overview.pdf},
  2022.

\bibitem{RTX4090}
NVIDIA.
\newblock Nvidia ada craft the engineering marvel of the rtx 4090.
\newblock
  \url{https://images.nvidia.com/aem-dam/Solutions/geforce/ada/ada-lovelace-architecture/nvidia-ada-gpu-craft.pdf},
  2022.

\bibitem{CUBLAS}
NVIDIA.
\newblock Basic linear algebra on nvidia gpus.
\newblock \url{https://developer.nvidia.com/cublas}, 2023.

\bibitem{Ada}
NVIDIA.
\newblock Nvidia ada gpu architecture.
\newblock
  \url{https://images.nvidia.com/aem-dam/Solutions/Data-Center/l4/nvidia-ada-gpu-architecture-whitepaper-v2.0.pdf},
  2023.

\bibitem{TensorCore}
NVIDIA.
\newblock Nvidia tensor cores unprecedented acceleration for hpc and ai.
\newblock \url{https://www.nvidia.com/en-us/data-center/tensor-cores/}, 2023.

\bibitem{GPT4}
OpenAI.
\newblock {GPT-4} technical report.
\newblock {\em CoRR}, abs/2303.08774, 2023.

\bibitem{iQAN}
Zhen Peng, Minjia Zhang, Kai Li, Ruoming Jin, and Bin Ren.
\newblock iqan: Fast and accurate vector search with efficient intra-query
  parallelism on multi-core architectures.
\newblock In Maryam~Mehri Dehnavi, Milind Kulkarni, and Sriram Krishnamoorthy,
  editors, {\em Proceedings of the 28th {ACM} {SIGPLAN} Annual Symposium on
  Principles and Practice of Parallel Programming, PPoPP 2023, Montreal, QC,
  Canada, 25 February 2023 - 1 March 2023}, pages 313--328. {ACM}, 2023.

\bibitem{ANN1999}
Sakti Pramanik and Jinhua Li.
\newblock Fast approximate search algorithm for nearest neighbor queries in
  high dimensions.
\newblock In Masaru Kitsuregawa, Michael~P. Papazoglou, and Calton Pu, editors,
  {\em Proceedings of the 15th International Conference on Data Engineering,
  Sydney, Australia, March 23-26, 1999}, page 251. {IEEE} Computer Society,
  1999.

\bibitem{PointNet}
Charles~Ruizhongtai Qi, Hao Su, Kaichun Mo, and Leonidas~J. Guibas.
\newblock Pointnet: Deep learning on point sets for 3d classification and
  segmentation.
\newblock In {\em 2017 {IEEE} Conference on Computer Vision and Pattern
  Recognition, {CVPR} 2017, Honolulu, HI, USA, July 21-26, 2017}, pages 77--85.
  {IEEE} Computer Society, 2017.

\bibitem{T5}
Colin Raffel, Noam Shazeer, Adam Roberts, Katherine Lee, Sharan Narang, Michael
  Matena, Yanqi Zhou, Wei Li, and Peter~J. Liu.
\newblock Exploring the limits of transfer learning with a unified text-to-text
  transformer.
\newblock {\em Journal of Machine Learning Research}, 21(140):1--67, 2020.

\bibitem{TTI1M}
Yandex Research.
\newblock Benchmarks for billion-scale similarity search.
\newblock
  \url{https://research.yandex.com/blog/benchmarks-for-billion-scale-similarity-search},
  2021.

\bibitem{L2IP2}
Anshumali Shrivastava and Ping Li.
\newblock Asymmetric {LSH} {(ALSH)} for sublinear time maximum inner product
  search {(MIPS)}.
\newblock In Zoubin Ghahramani, Max Welling, Corinna Cortes, Neil~D. Lawrence,
  and Kilian~Q. Weinberger, editors, {\em Advances in Neural Information
  Processing Systems 27: Annual Conference on Neural Information Processing
  Systems 2014, December 8-13 2014, Montreal, Quebec, Canada}, pages
  2321--2329, 2014.

\bibitem{DiskANN}
Harsha~Vardhan Simhadri, Ravishankar Krishnaswamy, Gopal Srinivasa,
  Suhas~Jayaram Subramanya, Andrija Antonijevic, Dax Pryce, David Kaczynski,
  Shane Williams, Siddarth Gollapudi, Varun Sivashankar, Neel Karia, Aditi
  Singh, Shikhar Jaiswal, Neelam Mahapatro, Philip Adams, Bryan Tower, and Yash
  Patel.

\bibitem{LLAMA7B}
Hugo Touvron, Thibaut Lavril, Gautier Izacard, Xavier Martinet, Marie{-}Anne
  Lachaux, Timoth{\'{e}}e Lacroix, Baptiste Rozi{\`{e}}re, Naman Goyal, Eric
  Hambro, Faisal Azhar, Aur{\'{e}}lien Rodriguez, Armand Joulin, Edouard Grave,
  and Guillaume Lample.
\newblock Llama: Open and efficient foundation language models.
\newblock {\em CoRR}, abs/2302.13971, 2023.

\bibitem{Attention}
Ashish Vaswani, Noam Shazeer, Niki Parmar, Jakob Uszkoreit, Llion Jones,
  Aidan~N. Gomez, Lukasz Kaiser, and Illia Polosukhin.
\newblock Attention is all you need.
\newblock In Isabelle Guyon, Ulrike von Luxburg, Samy Bengio, Hanna~M. Wallach,
  Rob Fergus, S.~V.~N. Vishwanathan, and Roman Garnett, editors, {\em Advances
  in Neural Information Processing Systems 30: Annual Conference on Neural
  Information Processing Systems 2017, December 4-9, 2017, Long Beach, CA,
  {USA}}, pages 5998--6008, 2017.

\bibitem{kdtree}
Ingo Wald and Vlastimil Havran.
\newblock On building fast kd-trees for ray tracing, and on doing that in o(n
  log n).
\newblock In {\em 2006 IEEE Symposium on Interactive Ray Tracing}, pages
  61--69, 2006.

\bibitem{wang2012scalable}
Jing Wang, Jingdong Wang, Gang Zeng, Zhuowen Tu, Rui Gan, and Shipeng Li.
\newblock Scalable k-nn graph construction for visual descriptors.
\newblock In {\em 2012 IEEE Conference on Computer Vision and Pattern
  Recognition}, pages 1106--1113. IEEE, 2012.

\bibitem{pami2005}
Liwei Wang, Yan Zhang, and Jufu Feng.
\newblock On the euclidean distance of images.
\newblock {\em {IEEE} Trans. Pattern Anal. Mach. Intell.}, 27(8):1334--1339,
  2005.

\bibitem{ANNSurvey}
Mengzhao Wang, Xiaoliang Xu, Qiang Yue, and Yuxiang Wang.
\newblock A comprehensive survey and experimental comparison of graph-based
  approximate nearest neighbor search.
\newblock {\em Proc. {VLDB} Endow.}, 14(11):1964--1978, 2021.

\bibitem{DCN}
Ruoxi Wang, Rakesh Shivanna, Derek~Zhiyuan Cheng, Sagar Jain, Dong Lin, Lichan
  Hong, and Ed~H. Chi.
\newblock {DCN} {V2:} improved deep {\&} cross network and practical lessons
  for web-scale learning to rank systems.
\newblock In Jure Leskovec, Marko Grobelnik, Marc Najork, Jie Tang, and Leila
  Zia, editors, {\em {WWW} '21: The Web Conference 2021, Virtual Event /
  Ljubljana, Slovenia, April 19-23, 2021}, pages 1785--1797. {ACM} / {IW3C2},
  2021.

\bibitem{PointConv}
Wenxuan Wu, Zhongang Qi, and Fuxin Li.
\newblock Pointconv: Deep convolutional networks on 3d point clouds.
\newblock In {\em {IEEE} Conference on Computer Vision and Pattern Recognition,
  {CVPR} 2019, Long Beach, CA, USA, June 16-20, 2019}, pages 9621--9630.
  Computer Vision Foundation / {IEEE}, 2019.

\bibitem{Tigris}
Tiancheng Xu, Boyuan Tian, and Yuhao Zhu.
\newblock Tigris: Architecture and algorithms for 3d perception in point
  clouds.
\newblock In {\em Proceedings of the 52nd Annual {IEEE/ACM} International
  Symposium on Microarchitecture, {MICRO} 2019, Columbus, OH, USA, October
  12-16, 2019}, pages 629--642. {ACM}, 2019.

\bibitem{zhang2018efficient}
Jialiang Zhang, Soroosh Khoram, and Jing Li.
\newblock Efficient large-scale approximate nearest neighbor search on opencl
  fpga.
\newblock In {\em Proceedings of the IEEE Conference on Computer Vision and
  Pattern Recognition}, pages 4924--4932, 2018.

\bibitem{SQ}
Wengang Zhou, Yijuan Lu, Houqiang Li, and Qi~Tian.
\newblock Scalar quantization for large scale image search.
\newblock In Noboru Babaguchi, Kiyoharu Aizawa, John~R. Smith, Shin'ichi Satoh,
  Thomas Plagemann, Xian{-}Sheng Hua, and Rong Yan, editors, {\em Proceedings
  of the 20th {ACM} Multimedia Conference, {MM} '12, Nara, Japan, October 29 -
  November 02, 2012}, pages 169--178. {ACM}, 2012.

\bibitem{RTNN}
Yuhao Zhu.
\newblock {RTNN:} accelerating neighbor search using hardware ray tracing.
\newblock In Jaejin Lee, Kunal Agrawal, and Michael~F. Spear, editors, {\em
  PPoPP '22: 27th {ACM} {SIGPLAN} Symposium on Principles and Practice of
  Parallel Programming, Seoul, Republic of Korea, April 2 - 6, 2022}, pages
  76--89. {ACM}, 2022.

\end{thebibliography}
\clearpage
\end{document}